\documentclass[runningheads]{llncs}
\usepackage[T1]{fontenc}
\usepackage{amsmath,amssymb,mathtools}
\usepackage{graphicx,booktabs,array}
\usepackage{tikz}
\usetikzlibrary{arrows.meta,positioning,fit,calc,decorations.pathreplacing,shapes.geometric,backgrounds}
\usepackage[protrusion=true,expansion=false]{microtype}
\usepackage{xurl}
\usepackage[hidelinks]{hyperref}
\newcommand{\R}{\mathbb R}
\newcommand{\E}{\mathbb E}
\newcommand{\one}{\mathbf 1}
\newcommand{\ADP}{\operatorname{ADP}}

\graphicspath{{figures/}}

\spnewtheorem{assumption}{Assumption}{\bfseries}{\itshape}
\spnewtheorem{prop}{Proposition}{\bfseries}{\itshape}
\spnewtheorem{pclaim}{Claim}{\bfseries}{\itshape}

\newcommand{\proofstep}[1]{\par\smallskip\noindent\textbf{#1}\enspace}



\newcommand{\RefinedWorking}{19.826087}
\newcommand{\RefinedJunior}{2.782609}
\newcommand{\RefinedTotal}{22.608696}
\newcommand{\RefinedDelta}{3.826087}

\newcommand{\ExactWorking}{19.404255}
\newcommand{\ExactDelta}{3.404255}
\newcommand{\ExactJunior}{2.723404}
\newcommand{\ExactTotal}{22.127660}
\newcommand{\ExactMinSavingIsolated}{1.179668}
\newcommand{\ExactMinSavingComplete}{1.472119}
\newcommand{\ExactMinSavingMoment}{0.024699}

\tikzset{
 user/.style={draw,rounded corners,fill=gray!8,minimum width=2.05cm,minimum height=.68cm,align=center},
 claim/.style={draw,fill=gray!3,minimum width=2.3cm,minimum height=.72cm,align=center},
 pool/.style={draw,double,double distance=.6pt,fill=gray!14,minimum width=2.25cm,minimum height=.72cm,align=center},
 box/.style={draw,rounded corners,align=center,minimum height=.72cm},
 flow/.style={-{Stealth[length=2mm]},thick},
 eligible/.style={dashed,gray!75},
 own/.style={thin},
 cut/.style={draw,dashed,rounded corners,inner sep=7pt,thick},
 note/.style={align=center,font=\footnotesize}
}

\title{Bottlenecked: The Capital Cost of Protected DeFi Lending}
\titlerunning{The Capital Cost of Protected DeFi Lending}
\author{Tarun Chitra\inst{1} \and Manvir Schneider\inst{2} \and Nagu Thogiti\inst{3}}
\authorrunning{T. Chitra et al.}
\institute{Gauntlet, \email{tarun@gauntlet.xyz} \and
Cardano Foundation, \email{manvir.schneider@cardanofoundation.org} \and
BreakwaterAI, \email{0xnagu@gmail.com}}
\hypersetup{pdftitle={Bottlenecked: The Capital Cost of Protected DeFi Lending},pdfauthor={Tarun Chitra, Manvir Schneider, Nagu Thogiti},pdfsubject={Research paper on protected DeFi lending}}
\begin{document}
\maketitle
\begin{abstract}
Securing hundreds of billions of dollars in decentralized finance (DeFi) requires both cryptographically secured collateral and economic capital to absorb losses that collateral alone cannot cover. Composability makes both the amount and location of this protection matter: Kelp's nearly \$300 million April 2026 bridge exploit showed how an attacker could borrow against stolen assets across lending markets. We study how to allocate junior capital that absorbs losses before users incur losses in lending protocols. The main object we study is a tripartite graph that records user assets, their covered liabilities, and the junior capital permitted to pay them. Cuts in this graph determine what protection is missing, while a static security game on the graph defines how much such protection should cost. We call this cost the adversarial default premium (ADP), which can be viewed as the excess cost of insurance in composable financial systems. We then study the joint cost of funding loans and protecting their creditors via this insurance. In a tripartite graph with $N$ borrowers and $M$ lenders, we show that existing systems such as Aave and Morpho have an insurance cost that is $\Theta(N)$. Under mild assumptions, we show that the optimal graph has an insurance cost that is $\Theta(\sqrt{N})$ with small constants. We apply our model to numerous 2026 DeFi cyber attacks and estimate ADP and the performance of optimized graphs. We find practical capital cost savings of up to 34.5\% for our optimized design versus designs that resemble Aave and Morpho. Our results suggest that new lending protocols can improve capital efficiency by treating junior capital as a first-class constraint in protocol design.
\end{abstract}

\section{Introduction}
An asset can lose its backing in one protocol while remaining acceptable collateral in another. In the April 2026 rsETH incident, a forged bridge message released 116,500 rsETH, some of which the attacker then supplied as collateral in lending markets. Aave service providers reported that their contracts continued to function as designed~\cite{KelpReport2026}. The receiving contracts had not removed the exposure to the failed bridge. Eventual losses depended on recovery and allocation decisions.
Measurements of deployed DeFi systems document nested contract interactions and shared financial building blocks~\cite{Kitzler2023}.

A curator choosing collateral permissions chooses two capital requirements. If either of two markets may request a dollar, but never both together, one shared dollar supplies the service that needs two dedicated dollars. Wider borrowing access releases idle working capital, but also lets a failed source reach more of it. Separately committed \emph{junior capital} pays covered losses before they fall on lending users. Sharing this protection helps settlement: the same funded pool can pay whichever account is hit. Our comparisons hold that payment right fixed.

We choose permissions and both commitments for a fixed lending service. A heterogeneous lower bound shows that excess working capital and worst-source exposure cannot both be small. Dispersed permissions attain square-root extra funding cost when demand and account counts grow proportionally and sufficient capacity remains. Isolation and complete access have linear overhead, from idle working capital and junior protection respectively. For a fixed graph, underwriting certificates test separate capital budgets. On a 57-account graph, a computer-assisted argument gives the exact minimum working and junior commitments, even allowing unequal balances.

The experiments test the design mechanism and measure the remaining opportunity in historical calibrations. Exact small-system optimization yields a 25\% improvement over isolation and complete access in one three-type case. Grouped Morpho calibrations offer smaller savings: concurrent debt requirements and concentrated sources leave isolation close to the capital floors. Selected designs mostly release working capital instead of reducing isolation's junior requirement. We also freeze the saved pre-event designs and check their service on later debt vectors.

Our baseline grants \emph{full-account rights}: a failed source can exhaust each permitted balance once before access closes. Persistent source-wide limits or account-local quotas can narrow these rights; inherited receipt permissions can enlarge them. The following section states the security model. Sections~\ref{sec:physical}--\ref{sec:underwriting} derive the capital bounds, Section~\ref{sec:continuation} treats additional access, and Section~\ref{sec:empirical} evaluates the designs and their limits.

\section{Security setting and lending model}
\label{sec:security}
Formal economic-security models examine adversarial extractable value and losses to designated victim contracts~\cite{Clockwork,BMZNoninterference2025,BMZFramework2026}. We study covered lender claims after a collateral-source failure; these need not equal attacker profit or a change in one contract's token holdings. The attacker controls that type's requests and chooses their accounts and order. Caller addresses do not define independent failures. We retain ledger execution, source assignment, permission enforcement, principal accounting, and availability of junior capital as trusted components.

Junior capital is committed before failure. Each account's principal can be lost once. Ordinary service is a separate promise to fund an observed batch; our upper construction does not establish online service after arbitrary irreversible earlier loans. An online policy providing the same service and attack scope still faces the lower bound.

Euler's configurable collateral-vault relationships provide candidate borrowing permissions, and EVC batching can execute a proposed allocation~\cite{EulerMarkets,EulerEVC}. Valid collateral allocation, final account checks, and a separate funded payment contract remain necessary. Appendix~\ref{app:protocol} gives the mapping; neither the model nor the empirical redesign is a deployed Euler implementation.

\subsection{Accounts and borrowing permissions}
\label{sec:model}
Fix a num\'eraire and a coverage period over which the capital remains committed. Let $[d]=\{1,\ldots,d\}$. We use $N$ borrowing types and $M$ lending accounts. Requests belong to the same type when they share a source and borrowing permissions. Assets exposed to a common failure must be grouped accordingly or included in a joint attack. An account may combine suppliers with identical rights, so neither $N$ nor $M$ counts individual suppliers. Grouping several correlated assets does not automatically preserve the unit request bound; the service set must be stated for the chosen groups.

Write $H\in\{0,1\}^{N\times M}$ for the Boolean borrowing-permission matrix. Account $f$ has a normalized upper capacity of one and commits $a_f\in[0,1]$ to lending. These commitments give working capital $C=\sum_fa_f$ and determine what can be lent or lost; unused capacity contributes to neither. The upper bound is part of the compared facility, not a claim that deployed vaults have equal balances. For an account set $S$, write $a(S)=\sum_{f\in S}a_f$. The accounts accessible to borrowing types $U\subseteq[N]$ are
\[
 N_H(U)=\{f\in[M]:\exists i\in U,\ H_{if}=1\},
 \qquad N_H(i)=N_H(\{i\}).
\]

\subsection{The same ordinary service in every comparison}
The facility promises every request
\begin{equation}
 \mathcal D_s=\{x\in[0,1]^N:\one^\top x\le s\},\qquad
 1\le s<N,\quad s\le M.
 \label{eq:demand}
\end{equation}
After observing $x$, a divisible allocation $z$ must satisfy
\begin{equation}
 z_{if}\ge0,\quad z_{if}=0\ (H_{if}=0),\quad
 \sum_fz_{if}=x_i,\quad\sum_i z_{if}\le a_f.
 \label{eq:allocation}
\end{equation}
The allocator observes the complete request before committing its loans against the available balances. An online guarantee would also have to account for earlier irreversible allocations. Recovered principal and subsequent request periods require their own service and coverage rules.

\begin{pclaim}[Exact service feasibility]
\label{cl:service}
Every request in $\mathcal D_s$ is serviceable if and only if
\begin{equation}
 a(N_H(U))\ge\min\{|U|,s\}\qquad\text{for every }U\subseteq[N].
 \label{eq:service}
\end{equation}
\end{pclaim}
\begin{proof}
For fixed $x$, connect an auxiliary source to type $i$ with capacity $x_i$, each type to its permitted accounts with a capacity larger than all finite money, and each account to a sink with capacity $a_f$. Flows are exactly the allocations in~\eqref{eq:allocation}. Max-flow/min-cut~\cite{FordFulkerson1956} gives missing service $\max_U[x(U)-a(N_H(U))]_+$. The largest permitted $x(U)$ is $\min\{|U|,s\}$, attained by unit requests and at most one fractional remainder. Enforcing~\eqref{eq:service} is therefore necessary and sufficient.
\end{proof}
Thus $C\ge s$, and each type reaches at least one working unit. The condition covers every admissible request, not only observed samples.

\subsection{The loss created by a permission}
Suppose one source fails while its type's full-account borrowing rights remain active. It may exhaust each accessible balance before access closes, with each account's principal spent at most once. We assign zero recovery to this covered stress and allow smaller draws as well. The unit bound on ordinary requests is not a cumulative attack limit. With one protected liability per account, the attainable pre-insurance claims and worst total loss are
\begin{equation}
 c_f^{\{i\}}=a_fH_{if},\qquad
 K^0(H,a)=\max_i a(N_H(i)).
 \label{eq:gross}
\end{equation}
Access is measured before an effective restriction takes effect. The rsETH report places the bridge release at 17:35 UTC and the start of reserve freezes around 19:00~\cite{KelpReport2026}. That response interval does not establish full exhaustion of any particular market. Flash loans can supply temporary capital for atomic attack sequences~\cite{FlashLoans}, but do not establish exhaustion of every permitted account before intervention. A pause that prevents it narrows the covered execution set.

The superscript zero denotes loss before outside protection. Different recoveries or failures affecting several types require the corresponding claim vectors. Section~\ref{sec:continuation} considers how later transactions can enlarge access.

We first allow every account liability to use one common pool of $J$ outside junior capital. The pool remains liquid and available under the covered failure. It pays up to the total claim, leaving $(\one^\top c-J)_+$ unpaid. Full protection therefore requires and is achieved by $J\ge K^0$. This reserve is additional to the account owners' working capital. For an implementation, it is sufficient to establish, for every covered execution $\omega$ from source $i$,
\begin{equation}
 0\le c_f^\omega\le a_fH_{if}\quad\text{for all }f.
 \label{eq:execution}
\end{equation}
Summing gives $\one^\top c^\omega\le K^0$, so $J\ge K^0$ pays the claims. Stricter concrete controls preserve this upper guarantee. Necessity uses the further assumption that each source can attain its full neighborhood loss.

When every funded account is reachable, dedicated protection must contain $a_f$ at account $f$: some source can exhaust it. Total junior capital then rises to $C$. Figure~\ref{fig:topologies} instead keeps a common pool in every panel. Appendix~\ref{app:payment} gives the payment cut for general eligibility and a bounded family of shared pools.

A tighter quantity limit changes the problem. If a source has an enforceable, non-replenishing allowance $\kappa_i$ over the coverage period, and may attain every draw up to that allowance within its neighborhood, common protection requires
\[
 J_{\kappa}(H,a)=\max_i\min\{\kappa_i,a(N_H(i))\}.
\]
An ordinary supply cap is not automatically such an allowance: its currency, scope, and reset rule matter. Under complete access, $a_f=s/M$ and $z_{if}=x_i/M$ serve every request in~\eqref{eq:demand}. A cumulative source cap of one, or persistent account-local limits $L_{if}=1/M$, also gives $J=1$. Both require source-preserving accounting across covered draws. The full-account frontier prices rights without these tighter limits.

\subsection{The endpoints and the benchmark}
With $M=N$, an isolated matching must hold $a_f=1$ at every account to serve a possible singleton unit request. Under common protection this gives $(C,J)=(N,1)$: little junior capital, but $N-s$ extra working dollars. Complete access can use $a_f=s/N$, giving $(C,J)=(s,s)$. It avoids the working surplus at the cost of exposing all $s$ dollars to one failed source. The sparse construction shares lending capacity while limiting that exposure.

For $N=M=7$ and $s=3$, Figure~\ref{fig:topologies} keeps the same common protection in every panel. Its sparse graph uses $C=7/2$ and $J=3/2$. Exact service cuts justify the half-unit balances; the general spectral certificate gives the larger amount $5/9$ (Appendix~\ref{app:seven}).
\newcommand{\TripartiteCaption}{Seven borrowing types and seven unit-capacity accounts provide the same service with common protection committed before failure. Account labels give $a_f$; diamonds give $J$. Red edges show one failed type's access; violet dashed edges are payment rights. The sparse allocation uses exact service cuts for this graph. Totals are capital quantities, not funding costs or global finite optima.}

\begin{figure}[t]
\centering
\begin{tikzpicture}[x=.94cm,y=1cm,
  src/.style={circle,draw=teal!80!black,fill=teal!8,minimum size=2.55mm,
    inner sep=0pt,line width=.45pt,font=\fontsize{6.6}{7}\selectfont},
  failed/.style={src,draw=red!65!black,fill=red!6},
  acct/.style={rectangle,rounded corners=1.5pt,draw=orange!80!black,
    fill=orange!8,minimum width=4.3mm,minimum height=2.65mm,
    inner sep=.6pt,line width=.45pt,font=\fontsize{7.1}{8}\selectfont},
  junior/.style={diamond,aspect=1.1,draw=violet!75!black,fill=violet!8,
    minimum width=6.7mm,minimum height=6.2mm,inner sep=.7pt,
    line width=.6pt,font=\fontsize{8.1}{9}\selectfont},
  access/.style={draw=black!23,line width=.27pt},
  attacked/.style={draw=red!65!black,line width=.75pt},
  cover/.style={draw=violet!27,densely dashed,line width=.35pt},
  usedcover/.style={draw=violet!75!black,densely dashed,line width=.55pt},
  labeltext/.style={align=center,font=\fontsize{7.1}{8}\selectfont,text=black!78,inner sep=0pt},
  paneltitle/.style={align=center,font=\fontsize{8.3}{9}\selectfont\bfseries,inner sep=0pt},
  amounts/.style={align=center,font=\fontsize{8.3}{9}\selectfont,inner sep=0pt}
]
\node[font=\small\bfseries] at (5.80,1.45) {Lending access and junior capital};
\node[font=\footnotesize,align=center] at (5.80,1.11)
 {Same service: $\mathcal D_3$. Same protection: one pool can pay every account.};
\foreach \panel/\offset/\heading/\bal in
 {a/.20/Isolated lending/1,b/4.35/Dispersed sparse lending/{1/2},c/8.50/Complete access/{3/7}}{
 \begin{scope}[shift={(\offset,0)}]
  \node[paneltitle] at (1.46,.71) {\heading};
  \node[labeltext] at (0,.35) {Types};
  \node[labeltext] at (1.32,.35) {Accounts};
  \node[labeltext] at (2.94,.35) {Junior\\capital};
  \foreach \i in {0,...,6}{
   \pgfmathtruncatemacro{\num}{\i+1}
   \ifnum\i=0
    \node[failed] (\panel L\i) at (0,-.37*\i) {\num};
   \else
    \node[src] (\panel L\i) at (0,-.37*\i) {\num};
   \fi
   \node[acct] (\panel M\i) at (1.32,-.37*\i) {$\bal$};
  }
 \end{scope}
}
\node[junior] (aR) at (3.14,-1.11) {$1$};
\node[junior] (bR) at (7.29,-1.11) {$3/2$};
\node[junior] (cR) at (11.44,-1.11) {$3$};
\foreach \xx in {3.14,7.29,11.44}{
 \node[labeltext] at (\xx,-1.77) {Prefunded};
}
\begin{scope}[on background layer]
 \foreach \i in {0,...,6}{
  \draw[access] (aL\i)--(aM\i);
  \foreach \offset in {0,1,3}{
   \pgfmathtruncatemacro{\j}{mod(\i+\offset,7)}
   \draw[access] (bL\i)--(bM\j);
  }
  \foreach \j in {0,...,6}{\draw[access] (cL\i)--(cM\j);}
  \draw[cover] (aR)--(aM\i);
  \draw[cover] (bR)--(bM\i);
  \draw[cover] (cR)--(cM\i);
 }
 \draw[attacked] (aL0)--(aM0);
 \draw[usedcover] (aR)--(aM0);
 \foreach \j in {0,1,3}{
  \draw[attacked] (bL0)--(bM\j);
  \draw[usedcover] (bR)--(bM\j);
 }
 \foreach \j in {0,...,6}{
  \draw[attacked] (cL0)--(cM\j);
  \draw[usedcover] (cR)--(cM\j);
 }
 \draw[black!12,line width=.35pt] (3.84,.70)--(3.84,-3.19);
 \draw[black!12,line width=.35pt] (7.99,.70)--(7.99,-3.19);
\end{scope}
\node[labeltext] at (1.66,-2.59) {One account exposed};
\node[labeltext] at (5.81,-2.59) {Three accounts exposed};
\node[labeltext] at (9.96,-2.59) {All accounts exposed};
\node[amounts] at (1.66,-2.90) {$C=7,\;J=1$};
\node[amounts] at (5.81,-2.90) {$C=7/2,\;J=3/2$};
\node[amounts] at (9.96,-2.90) {$C=3,\;J=3$};
\node[paneltitle] at (1.66,-3.18) {Total capital: 8};
\node[paneltitle] at (5.81,-3.18) {Total capital: 5};
\node[paneltitle] at (9.96,-3.18) {Total capital: 6};
\draw[attacked] (1.05,-3.60)--(1.48,-3.60);
\node[labeltext,anchor=west] at (1.60,-3.60) {Access from the failed type};
\draw[usedcover] (6.55,-3.60)--(6.98,-3.60);
\node[labeltext,anchor=west] at (7.10,-3.60) {Eligible junior payments};
\end{tikzpicture}
\caption{\TripartiteCaption}
\label{fig:topologies}
\end{figure}

At fixed permissions, write $M^0=\mathbb E_\mu\one^\top c^\omega$ for a benchmark mean and $\ADP^0=K^0-M^0$ for the pre-cover adversarial default premium. Full cover makes residual ADP zero. Our objective prices the working capital and protection needed for that cover; Appendix~\ref{app:benchmark} relates it to the static game.

\section{The service--exposure and funding frontier}
\label{sec:physical}
Let $\Delta=C-s$ be excess working liquidity. To keep each source's loss small, the facility must place substantial capital in accounts with few borrowing permissions. The service bottleneck is collective access: an allowed request can avoid a well-funded subset of those accounts, so enough capital must remain elsewhere to serve it. Proposition~\ref{prop:physical} quantifies the resulting surplus for unequal account balances.
\begin{prop}[Limited attack access requires extra lending liquidity]
\label{prop:physical}
Every serviceable graph in the model above obeys
\begin{equation}
 K^0\left(\Delta+\frac{C}{2M}\right)
 \ge\frac{C^2(N-s)}{4NM}
 \qquad\text{if }\Delta<C/2.
 \label{eq:physical}
\end{equation}
If $\Delta\ge C/2$, then $C\ge2s$ and $\Delta\ge s$.
\end{prop}

For an account set $S$, let $\Gamma_H(S)$ be the borrowing types with permission to some account in $S$. Account $f$ has degree $d_f=|\Gamma_H(\{f\})|$.
\begin{pclaim}[Stranded capital]
\label{cl:stranded}
If $|\Gamma_H(S)|\le N-s$, then $a(S)\le\Delta$.
\end{pclaim}
\begin{proof}
At least $\lceil s\rceil$ borrowing types cannot use any account in $S$. Place a permitted total request of $s$ on them, using $\lfloor s\rfloor$ unit requests and one fractional remainder when needed. Accounts outside $S$ must fund it, so $C-a(S)\ge s$.
\end{proof}

\begin{proof}[of Proposition~\ref{prop:physical}]
Count each working dollar once for each borrowing type allowed to reach it:
\[
 \sum_f a_fd_f=\sum_i\sum_fH_{if}a_f\le NK^0.
\]
Set $d_*=2NK^0/C$ and $L=\{f:d_f\le d_*\}$. Service makes $C,K^0>0$. If $a(L)<C/2$, then the accounts outside $L$ contribute more than $d_*C/2=NK^0$ to the preceding sum. Hence $a(L)\ge C/2$.

Let $k=\lfloor(N-s)/d_*\rfloor$. Every set $S\subseteq L$ with at most $k$ accounts has $|\Gamma_H(S)|\le kd_*\le N-s$, so Claim~\ref{cl:stranded} gives $a(S)\le\Delta$. If $k>0$ but $|L|<k$, apply this to $L$: it contradicts $a(L)\ge C/2>\Delta$. We can therefore choose the $k$ largest balances in $L$. Their total is at least $k a(L)/|L|\ge kC/(2M)$. For $k=0$, use the empty set and the same lower bound zero. In either case,
\[
 \Delta\ge\frac{kC}{2M}
 \ge\frac{C}{2M}\left(\frac{(N-s)C}{2NK^0}-1\right),
\]
where the second inequality uses $\lfloor z\rfloor\ge z-1$. Rearranging proves~\eqref{eq:physical}. Zero balances and unused accounts do not affect the counting argument.
\end{proof}

\subsection{Service from dispersed permissions}
Dispersed permissions let groups of types reach many distinct accounts. A classical spectral estimate~\cite{Tanner1984} bounds those neighborhoods; the account balances turn access into service.
\begin{pclaim}[A spectral witness for normal service]
\label{cl:spectral}
Let $W\in\mathbb Z_+^{N\times N}$ have row and column sums $d>0$, let $H=\one\{W>0\}$, and put $\sigma=\sigma_2(W)/d$. Assigning
\[
 a_*=\sigma^2+(1-\sigma^2)s/N
\]
to every account provides exact service and gives
\[
 C-s=\sigma^2(N-s),\qquad K^0\le da_*.
\]
The last inequality is an equality when $W$ is Boolean.
\end{pclaim}
The proof applies the second-singular-value bound and Cauchy--Schwarz to each neighborhood; Appendix~\ref{app:spectral-proof} gives the calculation.

For a $d$-regular bipartite Ramanujan multiplicity matrix, $\sigma^2\le4(d-1)/d^2$. Choose $J=K^0$. For $d\ge4$, Claim~\ref{cl:spectral} yields the degree-dependent construction bounds
\begin{equation}
 \boxed{\Delta\le\frac{4(N-s)}d,\qquad
 J\le d\frac{s}{N}+4\left(1-\frac{s}{N}\right).}
 \label{eq:degree-bounds}
\end{equation}
More accounts per type reduce the service surplus while increasing source exposure. For positive linear prices, the continuous degree minimizing this sufficient extra-cost certificate is
\begin{equation}
 d_{\rm cert}=2\sqrt{\frac{r_CN(N-s)}{r_Js}}.
 \label{eq:degree-rule}
\end{equation}
Round within the supported degree range. This selects a construction certificate, not the finite optimum over graphs. Appendix~\ref{app:degree} gives the derivation and the smaller bound retaining $(d-1)/d$.

The all-size result~\cite{MSS2018} supplies a suitable multiplicity matrix; degree $\lceil\sqrt N\rceil$ gives both corrections of order $\sqrt N$. Cohen gives a polynomial-time construction~\cite{Cohen2016}. Our artifact uses the explicit prime-field family below.

\subsection{An explicit design and unequal account counts}
For a projective plane of prime-power order $q$, take points as borrowing types and lines as accounts. There are $N=q^2+q+1$ of each; each point meets $q+1$ lines, and two points share one line. The Boolean incidence matrix satisfies
\[
 HH^\top=qI+\one\one^\top.
\]
Claim~\ref{cl:spectral} therefore gives
\begin{equation}
 a_* =\frac{q+s}{(q+1)^2},\qquad
 C-s=\frac{q(N-s)}{(q+1)^2},\qquad
 K^0=\frac{q+s}{q+1}.
 \label{eq:explicit}
\end{equation}
The artifact uses normalized triples and modular dot products for prime $q$. The cyclic seven-type graph in Figure~\ref{fig:topologies} is an equivalent labeling of the order-two incidence. Equation~\eqref{eq:explicit} gives a sufficient allocation at every service level; exact small-graph cuts may require less capital.

A square auxiliary construction can serve a rectangular facility without introducing fictitious capital. Divide the same interval $[0,C)$ into $N$ intervals $I_f$ of length $a_*=C/N$ and $M$ intervals $J_g$ of length $b=C/M$. Let $L_{fg}=|I_f\cap J_g|$, and give type $i$ permission to actual account $g$ whenever an eligible auxiliary interval overlaps it. Every auxiliary allocation transfers as
\begin{equation}
 z'_{ig}=\sum_f z_{if}\frac{L_{fg}}{a_*}.
 \label{eq:transfer}
\end{equation}
The overlap rows sum to $a_*$ and columns to $b$. Hence the transferred rows still pay $x_i$ and each actual column uses at most $b$. If an auxiliary type reaches at most $d$ accounts, each such interval overlaps at most $\lceil M/N\rceil+1$ actual intervals. Consequently,
\begin{equation}
 K^0_{\rm rect}\le\min\{C,\,bd(\lceil M/N\rceil+1)\}.
 \label{eq:rectangular}
\end{equation}
This construction requires $C\le M$, since actual account capacities are at most one. It preserves service but may enlarge attack access. Appendix~\ref{app:construction} gives the full overlap argument and exact witnesses.

Headroom cannot simply be omitted. If $M=s$ is an integer, service and the unit account bounds force $a_f=1$ at every account. Each account must be reachable by at least $N-s+1$ types: otherwise $s$ types could request one each while avoiding it, leaving only $s-1$ usable units. Averaging account degrees gives
\begin{equation}
 K^0\ge\frac{s(N-s+1)}N.
 \label{eq:noheadroom}
\end{equation}
Thus comparable $M$ and $N$ alone do not ensure square-root exposure at a fixed interior service fraction.

\subsection{Pricing the two capital requirements}
\label{sec:funding}
A depositor forgoes other uses of capital committed to lending. A junior supplier also bears covered losses before the protected users. Let $F_{N,M}(C,J)$ be the actual total funding cost in the stated num\'eraire over the coverage period, including any repricing of existing commitments. Its arguments are capital quantities, not annualized returns.

\begin{assumption}[Bounded incremental capital cost]
\label{ass:cost}
On the compared range $[s,M]\times[0,M]$, required quantities are available. Positive constants $\underline r_C,\underline r_J,\overline r_C,\overline r_J$, independent of $N$, bound the actual increments:
\begin{align}
 \underline r_Ch_C+\underline r_Jh_J
 &\le F_{N,M}(C+h_C,J+h_J)-F_{N,M}(C,J)\notag\\
 &\le\overline r_Ch_C+\overline r_Jh_J
 \label{eq:cost}
\end{align}
for nonnegative increments whose endpoints lie in that range.
\end{assumption}
A supplier cap may invalidate availability. A quote $y(X)$ that reprices all committed capital has marginal total cost $y(X)+Xy'(X)$; bounding $y$ alone does not establish~\eqref{eq:cost}. The reference $F_{N,M}(s,0)$ is the cost of unavoidable working liquidity, not a fully protected design or the benchmark mean.

\begin{prop}[The optimal cost of protected lending]
\label{prop:frontier}
Fix constants $0<\theta_-\le\theta_+<1$, $0<\mu_-\le\mu_+<\infty$, and $\delta>0$. Let $N,M,s$ satisfy
\[
 \theta_-\le s/N\le\theta_+,\qquad
 \mu_-\le M/N\le\mu_+,\qquad M-s\ge\delta N.
\]
For all sufficiently large $N$, among the Boolean lending graphs, heterogeneous $a\in[0,1]^M$ satisfying~\eqref{eq:service}, and common outside $K^0(H,a)\le J\le M$, Assumption~\ref{ass:cost} gives
\begin{equation}
 \min_{\substack{H,a,J\\K^0(H,a)\le J\le M}}F_{N,M}(C,J)
 =F_{N,M}(s,0)+\Theta(\sqrt N).
 \label{eq:frontier}
\end{equation}
Every optimum has $C-s=\Theta(\sqrt N)$ and $J=\Theta(\sqrt N)$. No convexity, concavity, or separability is required. For $M=N$, the isolated and complete-access benchmarks instead have linear extra funding cost.
\end{prop}
\begin{proof}
For every feasible design,
\begin{equation}
 \underline r_C\Delta+\underline r_JJ
 \le F_{N,M}(C,J)-F_{N,M}(s,0)
 \le\overline r_C\Delta+\overline r_JJ.
 \label{eq:costsandwich}
\end{equation}
Since $K^0\le C\le M$, the choice $J=K^0$ lies in the stated domain and is optimal for fixed $H,a$ by strict increase in $J$. If $\Delta\ge C/2$, then $\Delta\ge s\ge\theta_-N$, already giving a linear lower bound.

Otherwise $C<2s$. Put $b_0=\theta_+/\mu_-$ and $A_0=\theta_-^2(1-\theta_+)/(4\mu_+)>0$. Since $C/(2M)<s/M\le b_0$ and $C\ge s\ge\theta_-N$, Proposition~\ref{prop:physical} implies
\[
 K^0(\Delta+b_0)\ge\frac{C^2(N-s)}{4NM}\ge A_0N.
\]
Consequently the lower side of~\eqref{eq:costsandwich} is at least
\[
 \underline r_C\Delta+
 \frac{\underline r_JA_0N}{\Delta+b_0}
 \ge2\sqrt{\underline r_C\underline r_JA_0N}-\underline r_Cb_0
 =\Omega(\sqrt N).
\]
This is AM--GM after adding and subtracting $\underline r_Cb_0$.

For the upper bound,~\eqref{eq:degree-bounds} supplies a square auxiliary graph with $C=s+O(\sqrt N)$ and degree at most $\lceil\sqrt N\rceil$. The headroom condition makes $C\le M$ for large $N$. Equations~\eqref{eq:transfer}--\eqref{eq:rectangular} transfer it to actual accounts with $K^0=O(\sqrt N)$ because $M/N$ is bounded. Buy common $J=K^0$ and apply the upper side of~\eqref{eq:costsandwich}.

Bounded increments imply continuity of $F$: join any two points through their coordinatewise maximum and sum the coordinate bounds. For each of finitely many Boolean $H$, the service-feasible capacities form a closed subset of $[0,1]^M$, and $K^0$ is a continuous maximum of linear functions. The feasible minimum is attained after eliminating $J$. Its cost upper bound and positive lower prices force $\Delta,J=O(\sqrt N)$. The product bound then forces both quantities to be $\Omega(\sqrt N)$.

For $M=N$, the endpoint pairs $(N,1)$ and $(s,s)$ give linear overhead by~\eqref{eq:costsandwich}. An isolated one-to-one matching is not assumed feasible when $M<N$.
\end{proof}

Cheaper senior funding may require an internal junior attachment and run out at a supply cap. With quotes $.22$ and $.06$, attachment 40, and senior cap 20, a fund of 48 costs 9.28 at $(j,v)=(40,8)$. The relaxed split $(32,16)$ costs eight but is inadmissible. Its increments stay positive and bounded; Appendix~\ref{app:funding} derives the example.

\section{Underwriting a fixed permission graph}
\label{sec:underwriting}
Fix a Boolean $N\times N$ graph $H$ that is $d$-regular on both sides. Put $\theta=s/N$, $\sigma=\sigma_2(H)/d$, and let $J_H^*$ be the least junior capital over its serviceable balances.

\begin{pclaim}[Junior capital on a fixed regular graph]
\label{cl:fixed-junior}
Allowing unequal balances,
\begin{equation}
 \theta d\le J_H^*\le\theta d+(1-\theta)d\sigma^2.
 \label{eq:fixed-junior}
\end{equation}
The upper allocation is $a_f=\theta+(1-\theta)\sigma^2$, with $C_{\rm cert}=s+(N-s)\sigma^2$. At fixed total $C$, if the uniform allocation is serviceable, the minimum junior amount is exactly $dC/N$.
\end{pclaim}
\begin{proof}
Each working dollar is accessible to $d$ types, so $K^0\ge N^{-1}\sum_i(Ha)_i=dC/N\ge ds/N$. Claim~\ref{cl:spectral} supplies the uniform upper allocation. At fixed $C$, a serviceable uniform allocation has every source exposure equal to the lower bound $dC/N$.
\end{proof}
If $\sigma^2\le4/d$, the gap in~\eqref{eq:fixed-junior} is at most $4(1-\theta)$. The lower coefficient $d$ requires regularity of the Boolean permissions; it cannot be inferred from regularity of a multigraph whose support is irregular.

\paragraph{Testing two capital budgets.}
With lending budget $\bar C\le N$ and junior budget $\bar J$, the uniform certificate is affordable exactly when
\begin{equation}
 \boxed{\theta+(1-\theta)\sigma^2\le
 \min\{\bar C/N,\bar J/d\}.}
 \label{eq:underwriting-test}
\end{equation}
This sufficient test can reject budgets that a sharper certificate admits. Appendix~\ref{app:underwriting} gives the degree window and fixed-graph optimization.

\paragraph{A moment certificate and an exact allocation.}
For the order-seven incidence graph and $s=16$, write $m_f$ for the number of selected types incident to account $f$. For a set of $u$ types, the Gram identity gives $\sum_fm_f=8u$ and $\sum_fm_f^2=u(u+7)$. Since $(m-2)(m-3)\ge0$ for positive integer $m$,
\begin{equation}
 |N_H(U)|\ge\left\lceil\frac{u(33-u)}6\right\rceil.
 \label{eq:integer-neighbors}
\end{equation}
This \emph{moment certificate} supplies every service cut at $a_f=8/23$. A finite search sharpens its sole remaining obstruction at $u=16$.

\begin{pclaim}[Exact capital on the 57-account graph; computer-assisted]
\label{cl:exact57}
Fix the order-seven incidence permissions, $s=16$, and the full-account single-source model with common protection. Among all serviceable balances on this graph, including unequal balances,
\begin{equation}
 C_{H}^{\min}=\frac{912}{47},\qquad J_H^*=\frac{128}{47}.
 \label{eq:junior57}
\end{equation}
The uniform allocation $a_f=16/47$ attains both minima simultaneously.
\end{pclaim}
\begin{proof}[Argument and finite check]
If sixteen types reached at most 46 accounts, eleven avoided accounts would reach at most 41 types. Incidence symmetry turns this into an eleven-type obstruction. Moment identities force a line containing exactly three selected types; a projective transformation fixes those three and one off-line type. Exhausting the remaining seven choices among 48 points rules out the obstruction. Thus every sixteen-type set reaches at least 47 accounts. Together with~\eqref{eq:integer-neighbors} for smaller sets, this proves service at $16/47$.

For necessity, 57 explicit sixteen-type sets each reach 47 accounts, with every account appearing 47 times. Adding their service cuts gives $47C\ge912$; degree averaging gives $J\ge8C/57$. Appendix~\ref{app:exact57} proves the reduction and documents independent Python and C++ checks. These are fixed-graph minima, not optima over graphs.
\end{proof}

Budgets $\bar C=19.5$ and $\bar J=2.75$ admit the exact allocation but not the moment certificate. At linear relative price $\alpha=r_J/r_C$, the exact pair costs less than $(57,1)$ and $(16,16)$ for $10/39<\alpha<589/27$. Appendix~\ref{app:refined-prices} separates spectral, moment, and exact allocations and evaluates the same supplier quotes at all three. The minimum-working endpoint across graphs is in Appendix~\ref{app:minimum-working}.

\section{Changes in failure scope and subsequent access}
\label{sec:continuation}
A lending receipt may be accepted as collateral elsewhere, creating access that its originating type did not have. To study one such expansion, suppose a receipt from account $f$ can acquire the permissions of any type incident to $f$. Repetition can then reach the entire bipartite component. We count each covered principal once and require these transactions to be executable before intervention; a graph cycle alone does not establish either premise.

\begin{prop}[Controlled continuation and unrestricted reuse]
\label{prop:continuation}
Retain the service, funding, and common-protection assumptions of Proposition~\ref{prop:frontier}. Suppose every design still permits the original source loss, and the sparse constructions admit a feasible full-protection policy with total gross claims at most $\Gamma K^0$, for a size-independent $\Gamma$. If the controls' cost is included in the bill or at most $O(\sqrt N)$, optimal extra funding cost remains $\Theta(\sqrt N)$.

If instead one compromised source can exhaust every account in its connected permission component before any allowed intervention stops that reach, every design needs linear extra funding cost, and complete access attains that order.
\end{prop}
The original source attack preserves the lower bound, while the funded sparse policy gives the upper bound. Under unrestricted component reach, either all components have fewer than $s$ types, forcing $C\ge N$, or one contains at least $s$ types and requires $J\ge s$. Appendix~\ref{app:continuation} gives the proof and sufficient controls.

The positive case needs full protection funded through settlement, not a partial stabilizer. A cumulative inherited allowance or a fixed number of stages with bounded monetary branching can bound gross claims while preserving service. The empirical stress tests below reoptimize the finite design under expanded source coalitions and component reach.

For a uniform projective allocation $a_f=a$, the worst loss from $r\ge1$ jointly failed types is
\begin{equation}
 K_r^0=a\min\{N,1+qr\},\qquad K_0^0=0.
 \label{eq:multisource}
\end{equation}
Collinear types attain the bound. At the exact $a=16/47$, one, two, four, and eight failures require $128/47$, $240/47$, $464/47$, and $912/47$ of common junior capital, respectively. Eight failures expose all $\ExactWorking{}$ working units. These amounts price a larger failure set on this graph; the one-source optimality theorem does not cover that set.

\section{Evaluation: exact designs and historical calibration}
\label{sec:empirical}
The evaluation asks when joint design improves on both endpoints and how much opportunity remains in historically calibrated books. Synthetic cases have exact global certificates. The historical optimization uses weighted, grouped inputs and numerical bounds. It does not use the asymptotic theorem's normalized demand regime.

\subsection{Matched contracts and historical inputs}
For type caps $q_i$, every design must serve
\[
 \mathcal D(q,s)=\{x:0\le x_i\le q_i,\ \one^\top x\le s\},
 \qquad a(N_H(U))\ge\min\{\textstyle\sum_{i\in U}q_i,s\}\quad\forall U.
\]
The second condition follows by the same flow argument as Claim~\ref{cl:service}. Isolation grants each type its own account; complete access grants every type every account; the cost-selected design optimizes the graph and balances. All three designs in each comparison share the same service, account ceilings, source-failure scope, and common junior-payment rights. We minimize $C+\alpha J$ at the assumed price $\alpha=3$, reporting $C$ and $J$ separately. The price is assumed, not an observed insurance quote.

The historical inputs are archived Morpho API debt and supply histories~\cite{MorphoAPI}. Three retrospectively selected cohorts, associated with Usual, Resolv, and Kelp, contain 11, 32, and 15 markets respectively. Each has five modeled borrowing groups and five funding accounts. The recorded collateral categories specify joint stresses; they do not establish that all instruments in an asset class share an oracle or bridge failure. At each of 22 dates, days $-7$ through $+14$, a trailing 28-day window sets $q_i$ to the largest daily group debt and $s$ to the largest total debt. These debt-stock envelopes are a modeled service requirement, not measurements of new loan requests. The graph is redesigned independently at every date; overlapping windows are not independent observations. Appendix~\ref{app:empirical} records cohort membership, anchors, and acquisition coverage.

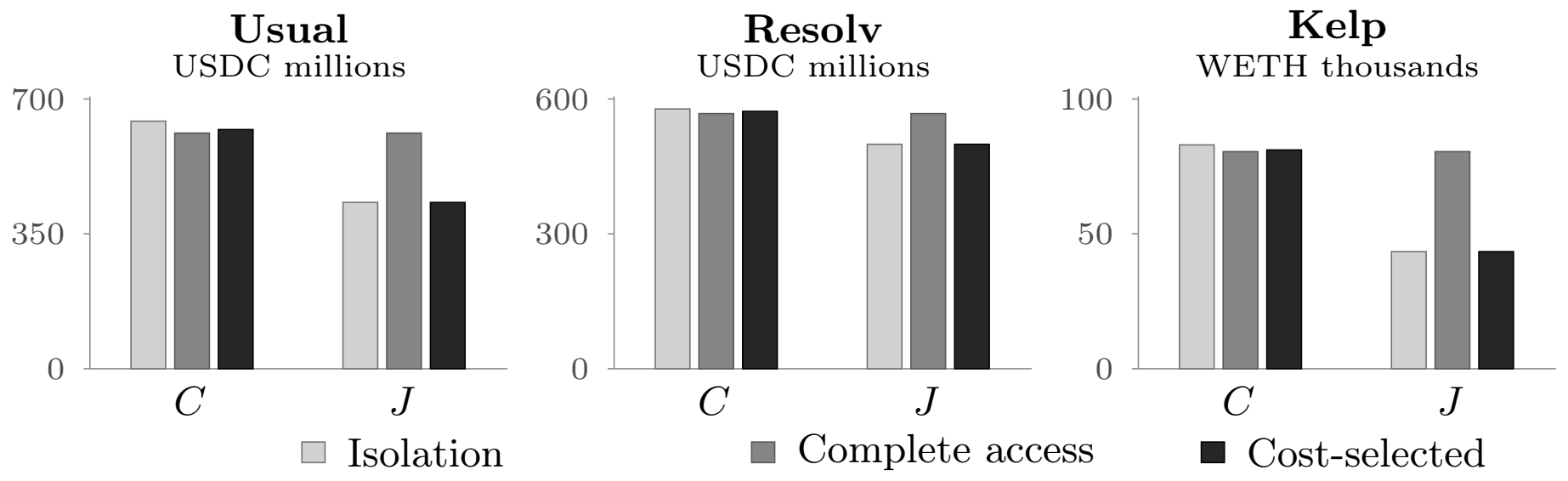
\begin{figure}[t]
\centering
\begin{tikzpicture}[x=1cm,y=1cm,font=\footnotesize]
\tikzset{ax/.style={draw=black!45,line width=.3pt}, tick/.style={font=\scriptsize,text=black!70}}
\begin{scope}[shift={(0.0,0)}]
\node[font=\small\bfseries] at (1.95,2.65) {Usual};
\node[font=\scriptsize] at (1.95,2.36) {USDC millions};
\draw[ax] (.40,0)--(.40,2.12); \draw[ax] (.40,0)--(3.65,0);
\draw[ax] (.35,0.000000)--(.40,0.000000); \node[tick,anchor=east] at (.32,0.000000) {0};
\draw[ax] (.35,1.050000)--(.40,1.050000); \node[tick,anchor=east] at (.32,1.050000) {350};
\draw[ax] (.35,2.100000)--(.40,2.100000); \node[tick,anchor=east] at (.32,2.100000) {700};
\filldraw[fill=black!18,draw=black!55,line width=.3pt] (0.720,0) rectangle (0.990,1.927353);
\filldraw[fill=black!48,draw=black!65,line width=.3pt] (1.060,0) rectangle (1.330,1.834638);
\filldraw[fill=black!85,draw=black,line width=.3pt] (1.400,0) rectangle (1.670,1.862097);
\node at (1.18,-.25) {$C$};
\filldraw[fill=black!18,draw=black!55,line width=.3pt] (2.370,0) rectangle (2.640,1.294745);
\filldraw[fill=black!48,draw=black!65,line width=.3pt] (2.710,0) rectangle (2.980,1.834638);
\filldraw[fill=black!85,draw=black,line width=.3pt] (3.050,0) rectangle (3.320,1.294745);
\node at (2.83,-.25) {$J$};
\end{scope}
\begin{scope}[shift={(4.08,0)}]
\node[font=\small\bfseries] at (1.95,2.65) {Resolv};
\node[font=\scriptsize] at (1.95,2.36) {USDC millions};
\draw[ax] (.40,0)--(.40,2.12); \draw[ax] (.40,0)--(3.65,0);
\draw[ax] (.35,0.000000)--(.40,0.000000); \node[tick,anchor=east] at (.32,0.000000) {0};
\draw[ax] (.35,1.050000)--(.40,1.050000); \node[tick,anchor=east] at (.32,1.050000) {300};
\draw[ax] (.35,2.100000)--(.40,2.100000); \node[tick,anchor=east] at (.32,2.100000) {600};
\filldraw[fill=black!18,draw=black!55,line width=.3pt] (0.720,0) rectangle (0.990,2.022418);
\filldraw[fill=black!48,draw=black!65,line width=.3pt] (1.060,0) rectangle (1.330,1.986437);
\filldraw[fill=black!85,draw=black,line width=.3pt] (1.400,0) rectangle (1.670,2.003607);
\node at (1.18,-.25) {$C$};
\filldraw[fill=black!18,draw=black!55,line width=.3pt] (2.370,0) rectangle (2.640,1.747326);
\filldraw[fill=black!48,draw=black!65,line width=.3pt] (2.710,0) rectangle (2.980,1.986437);
\filldraw[fill=black!85,draw=black,line width=.3pt] (3.050,0) rectangle (3.320,1.747326);
\node at (2.83,-.25) {$J$};
\end{scope}
\begin{scope}[shift={(8.16,0)}]
\node[font=\small\bfseries] at (1.95,2.65) {Kelp};
\node[font=\scriptsize] at (1.95,2.36) {WETH thousands};
\draw[ax] (.40,0)--(.40,2.12); \draw[ax] (.40,0)--(3.65,0);
\draw[ax] (.35,0.000000)--(.40,0.000000); \node[tick,anchor=east] at (.32,0.000000) {0};
\draw[ax] (.35,1.050000)--(.40,1.050000); \node[tick,anchor=east] at (.32,1.050000) {50};
\draw[ax] (.35,2.100000)--(.40,2.100000); \node[tick,anchor=east] at (.32,2.100000) {100};
\filldraw[fill=black!18,draw=black!55,line width=.3pt] (0.720,0) rectangle (0.990,1.742707);
\filldraw[fill=black!48,draw=black!65,line width=.3pt] (1.060,0) rectangle (1.330,1.690135);
\filldraw[fill=black!85,draw=black,line width=.3pt] (1.400,0) rectangle (1.670,1.703311);
\node at (1.18,-.25) {$C$};
\filldraw[fill=black!18,draw=black!55,line width=.3pt] (2.370,0) rectangle (2.640,0.911894);
\filldraw[fill=black!48,draw=black!65,line width=.3pt] (2.710,0) rectangle (2.980,1.690135);
\filldraw[fill=black!85,draw=black,line width=.3pt] (3.050,0) rectangle (3.320,0.911894);
\node at (2.83,-.25) {$J$};
\end{scope}
\filldraw[fill=black!18,draw=black!55,line width=.3pt] (2.05,-.73) rectangle (2.23,-.57);
\node[anchor=west] at (2.29,-.65) {Isolation};
\filldraw[fill=black!48,draw=black!65,line width=.3pt] (5.55,-.73) rectangle (5.73,-.57);
\node[anchor=west] at (5.79,-.65) {Complete access};
\filldraw[fill=black!85,draw=black,line width=.3pt] (9.05,-.73) rectangle (9.23,-.57);
\node[anchor=west] at (9.29,-.65) {Cost-selected};
\end{tikzpicture}
\caption{Required capital in the saved pre-event Morpho calibrations. Each cohort uses the same service set, repaired account ceilings, and $\alpha=3$ across designs. Working capital $C$ and junior capital $J$ are separate commitments in native units. These are counterfactual requirements, not observed protection balances.}
\label{fig:historical-snapshot}
\end{figure}

Observed group supply $b_i$ is raised to the common ceiling $\widetilde b_i=\max\{b_i,q_i\}$ so that the isolated comparator is feasible. The resulting commitments are charged in $C$, but their historical availability is assumed. Figure~\ref{fig:historical-snapshot} shows the saved day-$-1$ commitments. These are redesigns of grouped books, not deployed Morpho permission graphs. Morpho supplies the calibration data; the Euler discussion concerns a candidate implementation.

\subsection{Exact small-system optima and changed threat models}
For three types with $q_i=1$, $s=3/2$, and unit account ceilings, fund three accounts at $1/2$ and let each type reach two of them. Singletons reach one unit and every pair reaches all $3/2$. Thus $C=3/2$ and $J=1$. Every feasible design needs $C\ge3/2$ and $J\ge1$, so $C+3J=9/2$ is globally optimal. Isolation and complete access each cost six: the joint improvement is 25\%, with no reduction below isolation's junior requirement.

Exhaustive enumeration gives a 15\% saving over the cheaper endpoint for $q=(.8,1,1.2)$, $s=1.5$, and account ceilings 1.2; the four-unit-type case at $s=2$ gives 7.14\%. Across 28 deterministic cases, rational primal--dual certificates or infeasible service cuts cover all 7,116 canonical supports. These finite optima are distinct from the large-$N$ feasible construction.

The strict advantage disappears after reoptimization in the specified three-type tests with a two-source compromise, a hard cumulative loss cap of one, component-wide access, or the stronger imbalance $q=(.4,1,1.6)$ ($\beta=.6$). Coalition size is an attack budget, not a correlation estimate. The hard cap is an effective cumulative control available to all designs. Supply ceilings that make isolation infeasible are excluded from percentage comparisons against both endpoints. Appendix~\ref{app:exact-experiments} retains the complete grid and its negative results.

\subsection{Historically calibrated capital requirements}
The 66 five-group optimizations attain a maximum saved relative numerical gap of $5.05\times10^{-7}$. Table~\ref{tab:historical-ranges} gives the within-cohort range of joint-cost savings. Selected junior capital agrees with isolation on 58 dates and exceeds it on eight. The main gain comes from releasing working capital while preserving or occasionally increasing the worst-source protection amount. Complete access requires more junior capital at these prices; all comparators are funded for full protection under their modeled claims.

\begin{table}[t]
\centering\small
\setlength{\tabcolsep}{5pt}
\caption{Ranges across 22 rolling calibrations per Morpho cohort, at $\alpha=3$. Savings refer to $C+3J$, not $J$ alone. The last column counts dates on which selected junior capital exceeds isolation to the stated tolerance.}
\label{tab:historical-ranges}
\begin{tabular}{@{}lrrr@{}}
\toprule
Cohort & vs. isolation (\%) & vs. complete (\%) & Higher $J$\\
\midrule
Usual & 0.000--1.134 & 21.078--26.613 & 2/22 \\
Resolv & 0.213--0.507 & 8.200--10.211 & 6/22 \\
Kelp & 0.741--1.514 & 20.053--34.513 & 0/22 \\
 
\bottomrule
\end{tabular}
\end{table}

The small savings reflect a limited opportunity. In these fits $q_i\le s$. Put $Q=\sum_iq_i$ and $L=\max_iq_i$. Service forces $C\ge s$ and $J\ge L$, while isolation attains $(Q,L)$. Hence any design's fractional improvement over isolation is at most
\begin{equation}
 \frac{Q-s}{Q+\alpha L}.
 \label{eq:opportunity}
\end{equation}
The separate floors need not be jointly attainable. For Resolv, $s/Q$ is 97.85--98.45\% and the largest type holds 84.79--86.91\% of $Q$. The ceiling is only 0.438--0.606\%, against achieved savings of 0.213--0.507\%. Further graph search cannot yield a large gain under those fits.

The added capacity can be substantial. On Usual day $+12$, observed group supply is 169.571 million USDC, the ceiling adjustment adds 488.009 million, and selected working capital is 635.828 million. On Kelp day $+3$, added capacity is 27,407 WETH, 34.05\% of $s$. These plots therefore show required capital under the repaired model, not historically fundable deployments. Appendix~\ref{app:empirical} reports both added capacity and allocations above observed supply.

\subsection{Later demand against frozen permissions}
As an additional retrospective diagnostic, we freeze each saved day-$-1$ graph and its balances and test the fifteen subsequent daily group-debt vectors independently. We neither refit nor select a different tied optimum using later demands. Table~\ref{tab:frozen-service} compares the same frozen endpoint witnesses. In every failure a type cap is exceeded while the total remains below the frozen $s$. Additional access accommodates the changed composition in Resolv and Kelp; Usual retains the isolated failures.

\begin{table}[t]
\centering\small
\setlength{\tabcolsep}{3.8pt}
\caption{Independent later-demand feasibility tests, days 0--14. Counts use tolerance $10^{-6}$ in model units (one USDC or .001 WETH); shortfall magnitudes accompany the strict served-day criterion. Every complete-access shortfall is zero.}
\label{tab:frozen-service}
\begin{tabular}{@{}lccc l@{}}
\toprule
 & \multicolumn{3}{c}{Vectors fully served} & Maximum shortfall\\
Cohort & Selected & Isolated & Complete & Selected / isolated\\
\midrule
Usual & 7/15 & 7/15 & 15/15 & 15.129 / 15.129 m USDC \\
Resolv & 15/15 & 0/15 & 15/15 & 0 / 4.858 m USDC \\
Kelp & 15/15 & 0/15 & 15/15 & 0 / 988.59 WETH \\
 
\bottomrule
\end{tabular}
\end{table}

This test carries no loans, repayments, or depleted capital between days and does not establish an online lending guarantee. Cohorts and repaired ceilings remain retrospective. Several graphs can tie on the fitted objective. We evaluate the saved witness, not a uniquely selected deployment policy.

Minimizing the largest source loss also need not lower every source's loss. In the pre-event Kelp model, selected Kelp-group exposure rises from 496.19 to 37,743.09 WETH while the maximum stays at 43,423.54 WETH. The maximum-loss objective is indifferent to increases in nonbinding source exposures. These remain covered, but are not avoided Kelp losses. A secondary minimization of total exposure among equally priced designs would be a different selection rule; it is not used in these service tests. Likewise, lower pre-cover ADP can result entirely from a higher benchmark mean with unchanged $J$. Appendix~\ref{app:adp-diagnostic} retains both quantities.

The artifact separates exact synthetic certificates, numerical optimization on saved calibrations, and reconstruction from raw observations. All three cohorts' saved inputs can be checked; only Usual has the full raw input set in this release. No live transactions or user funds are used. The protocol obligations in Section~\ref{sec:security}, capital availability, and source grouping require separate validation before deployment.

\section{Related work and conclusion}
\label{sec:related}
Clockwork Finance formalizes extractable value in contract compositions~\cite{Clockwork}. MEV non-interference and local MEV address losses to designated victim contracts~\cite{BMZNoninterference2025,BMZFramework2026}; Priyadarshini and Bartoletti quantify amplification under component manipulation~\cite{Composition2025}. We choose borrowing permissions and funded protection together while preserving a fixed lending service.

Durvasula and Roughgarden bound cascading stake losses in restaking networks~\cite{RobustRestaking2025}. Chitra and Pai add service rewards and strategic operator rebalancing, and give an approximation algorithm for reward selection under their attack-profit assumptions~\cite{Restaking2024}. Our junior capital must pay the permitted lender claims without relying on incentives to exclude profitable attacks.

Bundi studies annual operational-loss tails, capital buffers, and supply-yield spreads~\cite{Bundi2026Tail}. Here protection covers an explicit execution set, while borrowing permissions determine exposure. The experimental relative funding price is assumed, not estimated from those spreads.

Manufacturing flexibility~\cite{JordanGraves1995} and queueing architectures~\cite{TsitsiklisXu2017} disperse access to reduce fragmentation. Here those same service edges expose accounts to a failed source. Our heterogeneous restriction couples the two quantities. Tanner's estimate and the Ramanujan constructions supply the spectral upper bound~\cite{Tanner1984,MSS2018,Cohen2016}; clearing, reinsurance, and convex-hull procurement supply the related payment and funding tools~\cite{EN2001,Reinsurance2020,HuaBaldick2017,UdellBoyd2016}.

On the fixed 57-account graph, the underwriting bounds coincide. Historical calibrations show why the structural opportunity need not produce large savings: debt requirements can already leave isolation close to the capital floors. Curators must evaluate service, source exposure, and fundable protection together before changing permissions.
\label{main-end}
\clearpage
\bibliographystyle{splncs04}
\bibliography{references}
\clearpage
\appendix
\renewcommand{\theHsection}{\Alph{section}}
\section{Payment rights and the common-capital specialization}
\label{app:payment}
A protected position identifies an owner, the account supplying principal, the recovery rule, accepted coverage, eligible junior capital, and settlement timing. A receipt is not automatically another insured liability. Let $Q\in\{0,1\}^{n\times m}$ assign each of $m$ liabilities to exactly one of $n$ users. For accepted user amounts $E\ge0$, every completed claim vector $c^\omega$ in a nonempty finite book $\Omega$ obeys $c^\omega\ge0$ and $Qc^\omega\le E$.

Let $A\in\{0,1\}^{p\times m}$ specify which of $p$ junior pools can pay each liability, and let $j\ge0$ be their committed available amounts. Payments satisfy $Z\ge0$, $Z_{fr}=0$ if $A_{fr}=0$, $Z^\top\one\le c$, and $Z\one\le j$. The minimum aggregate residual is
\[
 R_j(c)=\min_Z\one^\top(c-Z^\top\one).
\]
For $T\subseteq[m]$, let $N_A(T)=\{f:\exists r\in T,A_{fr}=1\}$.
\begin{pclaim}[The payment cut]
\label{prop:cuts}
The residual equals
\[
 R_j(c)=\max_{T\subseteq[m]}\{c(T)-j(N_A(T))\}.
\]
\end{pclaim}
\begin{proof}
Connect an auxiliary source to liability $r$ with capacity $c_r$, connect $r$ to every eligible pool with capacity greater than $\one^\top c+\one^\top j$, and connect pool $f$ to a sink with capacity $j_f$. A flow represents a feasible payment. A finite minimum cut with liabilities $T$ on the source side includes all their eligible pools and has value $c([m]\setminus T)+j(N_A(T))$. Max-flow/min-cut and subtraction from total claims give the formula. The empty set contributes zero.
\end{proof}

For a freely usable common pool, the formula reduces to $(\one^\top c-J)_+$. The source book in the main text has $Q=I_M$ and $c_f^{\{i\}}=a_fH_{if}$. Each such vector respects the accepted unit principal, and $J=K^0$ pays every vector. Loss of the junior assets, currency mismatch, or inability to pay by the contractual deadline requires a different supplied book or payment system.

\subsection{Restricted protection and procurement}
With existing junior capital $a^0$ and new commitments $k\ge0$, define $D(T)=\max_{\omega\in\Omega}c^\omega(T)$. Full protection is equivalent to
\[
 k(N_A(T))\ge[D(T)-a^0(N_A(T))]_+\quad(T\subseteq[m]).
\]
For fixed explicit claims and positive linear prices, minimizing $\kappa^\top k$ subject to these inequalities and actual supply caps is an LP. A feasible payment and a dual lower certificate establish its value. With implicit attack scenarios, finding a violated constraint remains a separate problem.

For example, suppose losses $(100,0)$ and $(0,100)$ are alternatives. Local commitments $k_A,k_B$ and common capital $g\le20$ satisfy $k_A+g\ge100$, $k_B+g\ge100$. Their total is at least $200-g\ge180$, attained by $(80,80,20)$. If both claims can occur together, full protection instead requires 200. Fixed eligibility and divisible capital make this placement calculation convex; the supplier attachment in Section~\ref{sec:funding} is a distinct procurement constraint.

Let $b(A)$ be the size of a supplied collection of junior pools covering all account liabilities. Suppose each has capacity to supply $K^0$. Assign each liability to one eligible selected pool, and put there the largest assigned total under a coherent source stress. Each amount is at most $K^0$ and no more than the sum of assigned individual maximum claims. Thus
\[
 K^0\le J_{\min}(A)\le\min\{C,b(A)K^0\}.
\]
A bounded covering family preserves the capital order if the actual commitments are available at the required prices. When every funded account is reachable, dedicated protection instead needs $J=C$, since each can lose its full $a_f$. The eligibility relation must authorize that sharing, even when the claims are statistically dependent.

\subsection{The benchmark and monetary units}
\label{app:benchmark}
For a fixed architecture, let $\Omega$ be its permitted completed stresses and $\mu$ a stated probability law on them. With common cover, put
\begin{align*}
 K(J)&=\max_{\omega\in\Omega}(\one^\top c^\omega-J)_+,\qquad
 M_\mu(J)=\mathbb E_\mu(\one^\top c^\omega-J)_+,\\
 \ADP_\mu(J)&=K(J)-M_\mu(J).
\end{align*}
ADP measures worst unpaid loss above the benchmark mean. At full cover, $J=K^0$ and residual ADP is zero. Before cover, write $M^0=\mathbb E_\mu\one^\top c^\omega$ and $\ADP^0=K^0-M^0$. For a common illustrative benchmark, let a one-unit default occur with probability $\pi$, with no claim otherwise. Singleton service makes that partial draw possible in every design, giving the same $M^0=\pi$. This chosen law allows a like-for-like comparison. Full protection still needs $K^0$ available capital; the expectation alone cannot pay a claim.
To state the static Stackelberg game, fix $j$ and $\mu$, and let $\mathcal Z_\omega$ be the feasible settlements of $c^\omega$. The defender commits to one rule $(Z^\omega)_{\omega\in\Omega}$ before the attacker selects $\omega$. Put $K(j)=\max_\omega R_j(c^\omega)$ and $M_\mu(j)=\mathbb E_\mu R_j(c^\omega)$. Then
\[
 \min_{(Z^\omega)\in\prod_\omega\mathcal Z_\omega}
 \max_{\omega\in\Omega}
 \{\one^\top(c^\omega-(Z^\omega)^\top\one)-M_\mu(j)\}
 =K(j)-M_\mu(j).
\]
Every feasible rule leaves at least $R_j(c^\omega)$ under each stress. Choosing a minimum-residual payment in each scenario attains all those minima, proving equality. The benchmark is fixed in this game. The earlier design decision instead prices the capital that makes every permitted claim payable; at full protection residual ADP is zero.

The common null/one-unit law above first selects a borrowing type by a declared distribution. In an active event of probability $\pi$, draw and lose one unit using a fixed order of eligible account indices; otherwise no claim occurs. The service promise makes this partial draw feasible for every architecture. Its location may change, but its total pre-cover mean remains $M^0=\pi$. Under linear prices,
\[
 (C-s)+\alpha(K^0-\pi)
\]
and normalized full overhead $(C-s)+\alpha K^0$ rank designs equally. At $\alpha=3$, $\pi=.1$, the values of the first expression for isolation, complete access, and the moment 57-account sparse design are $43.7$, $47.7$, and $2731/230\simeq11.873913$. The corresponding normalized full overheads are $44$, $48$, and $280/23\simeq12.173913$. The spectral pair gives $12.809375$ and $13.109375$, respectively. These differ by the common $\alpha\pi=.3$.

If the mean changes across designs, dropping it can reverse a comparison. For example, $(C,K^0,M^0,s)=(3,3,2.9,1)$ and $(2,2,0,1)$ at $\alpha=1$ give centered objectives $2.1$ and $3$, but full overheads $5$ and $3$. Funding amounts also differ from rates: converting a bill to a rate requires a stated coverage period and fee-paying base. These benchmark and price values are illustrative inputs.

\subsection{Price-weighted ADP interpretation}
Under linear prices, $F_{N,M}(C,J)=r_C(C+\alpha J)$, where $r_C>0$ and $\alpha=r_J/r_C>0$ is the junior-to-working price ratio; $\alpha\ge1$ when junior capital costs at least as much. With full protection $J=K^0$ and a common benchmark mean $M^0$,
\begin{equation}
 \frac{F(C,K^0)-F(s,0)}{r_C}
 =(C-s)+\alpha\ADP^0+\alpha M^0.
 \label{eq:weighted-objective}
\end{equation}
The last term is constant only when the same benchmark mean applies across the compared designs. Under that condition the choice minimizes a price-weighted sum of capital inefficiency and pre-cover ADP. The same argument does not justify minimizing residual ADP when the benchmark changes with the policy.

\section{Construction details and boundary cases}
\label{app:construction}
The singular-value argument and the heterogeneous lower bound are proved in the main text. The following details specify how the actual accounts are obtained and record finite boundary cases used by the artifact.

\subsection{Spectral service certificate}
\label{app:spectral-proof}
\begin{proof}[of Claim~\ref{cl:spectral}]
For $U$ of size $u>0$, let $y=W^\top\one_U$. Its support is $N_H(U)$ and $\one^\top y=du$. Decompose $\one_U=(u/N)\one+z$ with $z\perp\one$. Since $\|z\|_2^2=u(1-u/N)$, regularity and the second singular value give
\[
 \|y\|_2^2\le d^2u^2/N+d^2\sigma^2u(1-u/N).
\]
Cauchy--Schwarz on the support yields
\begin{equation}
 |N_H(U)|\ge\frac{u}{\sigma^2+(1-\sigma^2)u/N}.
 \label{eq:spectral-neighborhood}
\end{equation}
Here $0\le\sigma\le1$. For $u\le s$ the denominator is at most $a_*$. For $u\ge s$ the ratio is nondecreasing in $u$ and is at least $s/a_*$. Both give~\eqref{eq:service}. An integer row of total $d$ has at most $d$ distinct positive entries, proving the exposure bound. Multiplicities certify dispersion; they never duplicate account capital.
\end{proof}

\subsection{The explicit incidence family}
For a prime power $q$, projective points are one-dimensional subspaces of the three-dimensional vector space over the field of order $q$. Lines are the corresponding two-dimensional subspaces, represented dually by nonzero coefficient triples. Incidence is the vanishing of the dot product. There are $q^2+q+1$ points and lines, each point lies on $q+1$ lines, and each two points share exactly one. Hence $HH^\top=qI+\one\one^\top$. Its leading squared singular value is $(q+1)^2$ and all remaining squared singular values are $q$, proving~\eqref{eq:explicit}.

For $q=2$, the triples $\{i,i+1,i+3\}$ modulo seven specify the same incidence after relabeling. At $s=2$, $a_*=4/9$, $C=28/9$, and $K^0=4/3$. For $q=7,s=16$, the spectral allocation is $a_*=23/64$, $C=1311/64$, and $K^0=23/8$. Section~\ref{sec:underwriting} improves it to $a_f=8/23$ on the same graph. The artifact constructs prime fields using normalized triples and modular arithmetic and verifies the full integer Gram identity.

The all-size upper argument uses the integer biadjacency from the cited Ramanujan result. Collapsing multiplicities preserves support neighborhoods, though not necessarily Boolean regularity. Claim~\ref{cl:spectral} uses the original multiplicity matrix as a certificate and bounds each Boolean row's distinct degree by $d$. The implementation covers the explicit prime family, not the cited all-size construction algorithm.

\subsection{Capacity-preserving account reassignment}
\begin{pclaim}[Capacity-preserving account reassignment]
\label{cl:rectangular}
Suppose a Boolean square graph $H\in\{0,1\}^{N\times N}$ provides the required service with the same positive amount $a_*$ in every auxiliary account, and every borrowing type reaches at most $d$ such accounts. Let $C=Na_*$ and choose an integer $M$ with $C\le M$. There is a Boolean $N\times M$ permission graph with actual account balances $b=C/M\le1$ that provides the same service and obeys
\[
 K^0_{\rm rect}\le \min\{C,\,bd(\lceil M/N\rceil+1)\}.
\]
\end{pclaim}
\begin{proof}
\proofstep{Step 1: divide one capital line two ways.}
Partition $[0,C)$ into the $N$ half-open intervals
\[
 I_f=[(f-1)a_*,fa_*),\qquad f\in[N].
\]
These label the auxiliary balances. Partition the same line into the $M$ intervals $J_g=[(g-1)b,gb)$, $g\in[M]$, labeling the actual accounts. Let
\[
 L_{fg}=|I_f\cap J_g|.
\]
Row sums of $L$ are $a_*$ and column sums are $b$. Both systems contain exactly $C$; no overlap is counted as new money.

\proofstep{Step 2: transfer permissions and payments.}
Set $H'_{ig}=1$ precisely when some $f$ has $H_{if}=1$ and $L_{fg}>0$. Take any feasible auxiliary service flow $z_{if}$ for any permitted demand $x$. Define the actual flow
\[
 z'_{ig}=\sum_f z_{if}\frac{L_{fg}}{a_*}.
\]
Its row sums are $x_i$, because $\sum_gL_{fg}/a_*=1$. It uses only allowed actual permissions. For each actual account,
\[
 \sum_i z'_{ig}
 =\sum_f\frac{L_{fg}}{a_*}\sum_i z_{if}
 \le\sum_fL_{fg}=b.
\]
Thus every auxiliary feasible demand is actually funded within the physical balances.

\proofstep{Step 3: bound newly accessible capital.}
An interval of length $a_*$ intersects at most $\lceil a_*/b\rceil+1=\lceil M/N\rceil+1$ of the actual intervals with positive length. Endpoint contacts of length zero create no permission. Each type reaches at most $d$ auxiliary intervals, hence at most $d(\lceil M/N\rceil+1)$ actual accounts. Their total available capital is at most $b$ times that count and at most $C$. This bounds the actual saturated source claim. The construction may increase reachable capital relative to the square design, so we use this bound and do not assert exact preservation of attack loss.
\end{proof}

For $M/N$ bounded above and away from zero, $s/N$ bounded away from zero and one, and $M-s\ge\delta N>0$, the square spectral construction has $C=s+O(\sqrt N)\le M$ for all sufficiently large $N$. Its account reassignment has $K^0=O(\sqrt N)$, since $b\le1$, $d=O(\sqrt N)$, and $\lceil M/N\rceil+1$ is bounded. The construction is explicit given the square graph.

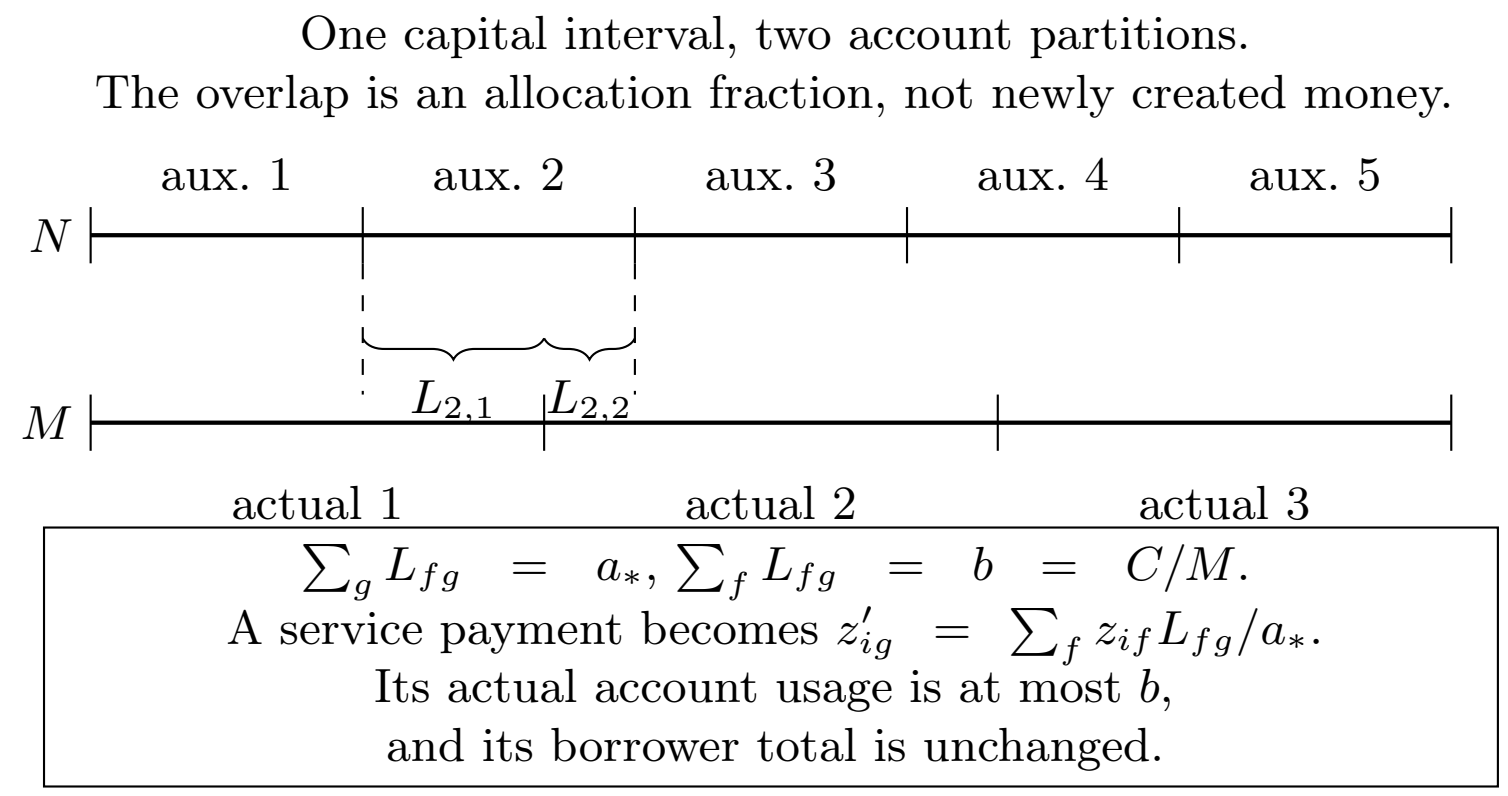
\begin{figure}[htbp]
\centering
\begin{tikzpicture}[x=.9cm,y=.62cm,>=Stealth,font=\footnotesize]
\node[align=center,text width=10cm] at (5,4.8) {One capital interval, two account partitions.\\The overlap is an allocation fraction, not newly created money.};
\draw[thick] (0,3)--(10,3);
\foreach \x in {0,2,4,6,8,10}{\draw (\x,2.7)--(\x,3.3);}
\foreach \x/\i in {1/1,3/2,5/3,7/4,9/5}{\node at (\x,3.65){aux. $\i$};}
\node[anchor=east] at (0,3){$N$};
\draw[thick] (0,1)--(10,1);
\foreach \x in {0,3.333,6.667,10}{\draw (\x,.7)--(\x,1.3);}
\foreach \x/\i in {1.667/1,5/2,8.333/3}{\node at (\x,.15){actual $\i$};}
\node[anchor=east] at (0,1){$M$};
\draw[dashed] (2,2.7)--(2,1.3);\draw[dashed] (4,2.7)--(4,1.3);
\draw[decorate,decoration={brace,mirror,amplitude=4pt}] (2,1.9)--node[below=.16cm] {$L_{2,1}$}(3.333,1.9);
\draw[decorate,decoration={brace,mirror,amplitude=4pt}] (3.333,1.9)--node[below=.16cm] {$L_{2,2}$}(4,1.9);
\node[draw,align=center,text width=9.4cm] at (5,-1.5) {$\sum_gL_{fg}=a_*$, $\sum_fL_{fg}=b=C/M$.\\A service payment becomes $z'_{ig}=\sum_f z_{if}L_{fg}/a_*$.\\Its actual account usage is at most $b$, and its borrower total is unchanged.};
\end{tikzpicture}
\caption{One capital interval partitioned into auxiliary and actual accounts. Positive overlap transfers borrowing permissions and the corresponding fraction of each service flow. Both partitions contain $C$ working units; account reassignment can broaden attack access, which is bounded separately.}
\label{fig:rectangular}
\end{figure}

\subsection{Why headroom is required}
At $M=s\in\mathbb N$, service gives $C\ge s=M$, while account capacities give $C\le M$. All actual balances therefore equal one. If an account were accessible to at most $N-s$ types, at least $s$ types could request one without access to it; only $s-1$ working units would remain. Every account has degree at least $N-s+1$. Summing those degrees and averaging over types proves~\eqref{eq:noheadroom}. This is a genuine failure of a square-root conclusion from comparable counts alone; it does not classify every sequence with shrinking positive headroom.

\subsection{Multiple compromised types}
For the projective family, a first point reaches $q+1$ lines. Each additional point shares at least one line with the first and adds at most $q$, so $r$ points reach at most $\min\{N,1+qr\}$ lines. Up to $q+1$ collinear points attain that bound: their common line is the only repeated line, and all their other incident lines are distinct. All points of one line reach every line. Multiplication by the common account amount $a$ proves~\eqref{eq:multisource}. With $N=57,s=16$, the spectral allocation gives losses $2.875$, $5.390625$, $10.421875$, and $20.484375$ for $r=1,2,4,8$. The moment $a=8/23$ gives $64/23$, $120/23$, $232/23$, and $456/23$. A source grouping affected by a common oracle or collateral failure must use the appropriate larger set.

\subsection{Allowing service shortfall}
If service may leave at most $\nu$ units unfunded in any request, the cut becomes
\[
 a(N_H(U))\ge\min\{|U|,s\}-\nu.
\]
The stranded-capital argument then gives $a(S)\le\Delta+\nu$. Whenever $\Delta+\nu<C/2$, the same counting and floor steps yield
\[
 K^0\left(\Delta+\nu+\frac C{2M}\right)
 \ge\frac{C^2(N-s)}{4NM}.
\]
Refused service is not an insured default loss without an additional contract. The main optimum concerns $\nu=0$.

\subsection{Exact seven-type comparison and a sequential-routing failure}
\label{app:seven}
Label types and accounts by $0,\ldots,6$ and use $N_H(i)=\{i,i+1,i+3\}$ modulo seven. For type-set cardinalities $u=1,\ldots,7$, the minimum neighborhood sizes are respectively
\[
 3,\;5,\;6,\;6,\;7,\;7,\;7.
\]
The first two follow from degree three and pairwise intersection one. A set of at least three types reaches at least six accounts: otherwise two excluded accounts would leave at most two eligible types outside their union of five neighbors. Four types can avoid one account by choosing its complementary four types, and each of the indicated bounds is attained. Five types cannot avoid any account, since its complement has only four types. Thus at $s=3$ the largest ratio $\min\{u,3\}/|N_H(U)|$ is $1/2$. This is the minimum uniform amount for this fixed graph. It gives $C=7/2$, $K^0=3/2$, and common full capital $C+J=5$. The isolated and complete-access pairs are $(7,1)$ and $(3,3)$, giving quantities eight and six. These are not funding costs or a proof of optimality over all seven-type graphs.

The general spectral sufficient amount at $q=2,s=3$ is $5/9$. Using exact cuts for Figure~\ref{fig:topologies} does not change equation~\eqref{eq:explicit}, nor the separate $s=2$ example. The artifact enumerates all subsets and supplies exact allocations for the 35 size-three extreme requests; the remaining points of $\mathcal D_3$ follow by divisibility and convexity.

Static feasibility does not validate arbitrary sequential splitting. In the $q=7,s=16$ design, let the eight points of one projective line each request one unit. Splitting each request equally among its eight incident accounts assigns one unit in total to their common account, exceeding both the spectral $23/64$ balance and the moment $8/23$ balance. The request totals only eight, below sixteen; a feasible jointly chosen allocation still exists. This is a counterexample to that routing rule, not to the service cut.

\subsection{Degree certificates and their price-dependent selection}
\label{app:degree}
For $d\ge4$, the Ramanujan witness has $\sigma^2\le4(d-1)/d^2\le1$. Before simplifying the factor $(d-1)/d$, Claim~\ref{cl:spectral} gives
\[
 \Delta\le\frac{4(d-1)(N-s)}{d^2},\qquad
 J=K^0\le d\frac{s}{N}+\frac{4(d-1)}d\left(1-\frac{s}{N}\right).
\]
Dropping $(d-1)/d\le1$ recovers~\eqref{eq:degree-bounds}. The two upper bounds refer to the constructed design with $J=K^0$, not every feasible reserve choice.

For positive linear prices put $U(d)=4r_C(N-s)/d+r_Jds/N+4r_J(1-s/N)$. Differentiating this coarse certificate gives
\[
 U'(d)=-\frac{4r_C(N-s)}{d^2}+\frac{r_Js}{N},\qquad
 U''(d)=\frac{8r_C(N-s)}{d^3}>0.
\]
Its unique unconstrained minimum is~\eqref{eq:degree-rule}. On an integer interval $d_{\min}\le d\le d_{\max}$, clamp that value to the interval and compare its neighboring integers. When availability restricts the degree set further, compare only supported certificates. Choosing the minimizer of the sharpened expression above instead is a separate one-dimensional comparison; it still minimizes an upper certificate, not the actual finite architecture cost. For nonlinear funding the upper incremental prices may replace $r_C,r_J$ to obtain a conservative design bound.

At fixed $s/N$ and fixed positive price ratios, both degree choices have square-root order. The general heterogeneous lower bound is what makes that order optimal. If an \emph{extra-funding} budget grows as $B$ while unavoidable working funding is supplied separately, inverting the frontier gives supported scale of order $B^2$. A junior-only budget gives no such comparison with isolation, whose common junior requirement is one.

For the all-size choice $d=\lceil\sqrt N\rceil\ge4$, the bounds used in Proposition~\ref{prop:frontier} are
\begin{equation}
 \Delta\le4(N-s)/d=O(\sqrt N),\qquad
 K^0\le ds/N+4(1-s/N)=O(\sqrt N).
 \label{eq:spectral-orders}
\end{equation}

\section{Underwriting certificates and the capital curve}
\label{app:underwriting}
This appendix uses the square model $M=N$, unit account capacities, the full-account single-source loss in~\eqref{eq:gross}, and common junior capital. Its fixed-graph statements optimize balances without changing permissions. Its capital curve permits the graph to change as well.

\subsection{A fixed graph and two budgets}
For a Boolean graph $H$ that is $d$-regular on both sides, each working dollar contributes to exactly $d$ source exposures. Thus
\[
 K^0(H,a)\ge\frac1N\one^\top Ha=\frac{dC}{N}.
\]
Together with the spectral service allocation, this proves Claim~\ref{cl:fixed-junior}. When $\sigma^2\le4/d$, it gives
\[
 \theta d\le J_H^*\le\theta d+4(1-\theta),\qquad
 \frac{J_{\rm cert}}{J_H^*}\le1+\frac{4(1-\theta)}{\theta d}.
\]
The gap is additive in normalized capital units. At $d=\Theta(\sqrt N)$ and a fixed interior service fraction, the relative gap is $O(N^{-1/2})$ on the fixed graph. Neither statement transfers the Boolean degree to an irregular support obtained by collapsing parallel edges of a regular multiplicity matrix.

For budgets $s\le\bar C\le N$ and $\bar J\ge0$, let $\bar\Delta=\bar C-s$. The test in~\eqref{eq:underwriting-test} is equivalently
\[
 \sigma^2\le\min\left\{
 \frac{\bar\Delta}{N-s},\frac{\bar J/d-\theta}{1-\theta}\right\}.
\]
It certifies the explicit uniform allocation in Claim~\ref{cl:fixed-junior}. Substituting $4/d$ for $\sigma^2$, when the candidate meets that spectral bound, gives the coarser sufficient window
\[
 \left\lceil\frac{4(N-s)}{\bar\Delta}\right\rceil
 \le d\le
 \left\lfloor\frac{\bar J-4(1-\theta)}{\theta}\right\rfloor,
 \qquad\bar\Delta>0.
\]
A degree in this window still needs a graph certificate. An empty window rules out this sufficient sizing rule, not every allocation within the budgets.

For an arbitrary fixed $H$, the direct optimization is
\begin{align*}
 \min_{a,J}\quad &J\\
 \text{subject to}\quad &a(N_H(U))\ge\min\{|U|,s\}\quad(U\subseteq[N]),\\
 &\one^\top a\le\bar C,\qquad Ha\le J\one,\qquad 0\le a\le1,\quad J\ge0.
\end{align*}
A feasible allocation and a lower certificate bracket its value. Feasibility requires satisfying every service cut, and there may be exponentially many. Sampled requests alone cannot supply the upper certificate. The integer-moment argument and explicit cut orbit below avoid exhaustive enumeration at $N=57$.

\subsection{The 57-account moment certificate}
\label{app:moment57}
The order-seven incidence has $N=57$, degree eight, and $HH^\top=7I+\one\one^\top$. For $U$ of size $u$, set $m=H^\top\one_U$. Then
\[
 \sum_fm_f=8u,\qquad\sum_fm_f^2=u(u+7).
\]
Every positive integer $m_f$ obeys $m_f^2\ge5m_f-6$. Summing over its support proves~\eqref{eq:integer-neighbors}. At sizes $u=1,\ldots,16$, the integer lower bounds on the neighborhood size are
\[
 6,11,15,20,24,27,31,34,36,39,41,42,44,45,45,46.
\]
Each supplies at least $u$ when multiplied by $8/23$. Sets of more than sixteen types contain a sixteen-type subset. Therefore all service cuts hold at $a_f=8/23$, giving
\[
 C=\frac{456}{23},\qquad K^0=J=\frac{64}{23},\qquad C+J=\frac{520}{23}.
\]
The spectral certificate uses $a_f=23/64$. The smaller moment allocation uses integrality of the incidence counts, not a sharper general spectral estimate.

\subsection{An explicit 57-cut lower certificate}
\label{app:orbit57}
Index the normalized nonzero triples over $\mathbb F_7$ lexicographically, with the first nonzero coordinate equal to one. Use these representatives for both rows and columns, with $H_{if}=1$ when the dot product is zero. In zero-based row indices, take
\[
 U=\{1,3,4,10,14,25,26,29,30,31,37,38,41,42,53,55\}.
\]
Its neighborhood contains 47 accounts. The finite certificate records that neighborhood and all images under the invertible coordinate map
\[
 G=\begin{pmatrix}0&0&6\\1&0&1\\0&5&4\end{pmatrix}
 \quad\text{over }\mathbb F_7.
\]
For row-vector coordinates, transform types by $p\mapsto pG$ and accounts by $\ell\mapsto\ell(G^{-1})^\top$, normalizing afterward. These transformations preserve dot products, hence permissions. The artifact gives the inverse, the two induced permutations, and the 57 transforms of $U$.

Direct integer checks establish that each transformed set has sixteen types and 47 neighbors, and that every account belongs to exactly 47 of those neighborhoods. Every set carries a permitted sixteen-unit request. Adding its service inequality to the other 56 yields
\[
 47C\ge57\cdot16=912.
\]
Since the graph is degree eight, $57J\ge8C$, whence $J\ge128/47$. Together with the moment allocation,
\[
 2.723404\ldots=\frac{128}{47}\le J_H^*
 \le\frac{64}{23}=2.782608\ldots,
 \qquad \frac{J_{\rm cert}}{J_H^*}\le\frac{47}{46}.
\]
The witness establishes a necessary bound for all serviceable balances on this graph, including unequal balances. The witness alone does not establish that 47 is the smallest neighborhood. It was found using a local search. Verifying it and the lower bound requires only the listed integer data. Appendix~\ref{app:exact57} proves the matching neighborhood bound by exhaustive search. At fixed $C=456/23$, the upper allocation attains the averaging lower bound exactly.

For budgets $\bar C=20$ and $\bar J=14/5$, the moment allocation fits both. The spectral allocation $C=1311/64$, $J=23/8$ exceeds both. The budget test therefore benefits from a stronger service certificate without changing the graph or adding capital rights.

\subsection{Junior capital as a function of working capital}
\label{app:capital-curve}
Define the best junior amount at exactly $C\in[s,N]$ working capital by
\[
 \mathcal J_{N,s}(C)=\min_{H,a}\left\{\max_i(Ha)_i:
 \begin{array}{l}
 H\in\{0,1\}^{N\times N},\ 0\le a\le1,\ \one^\top a=C,\\
 a(N_H(U))\ge\min\{|U|,s\}\quad(U\subseteq[N])
 \end{array}\right\}.
\]
At $\Delta=C-s<C/2$, Proposition~\ref{prop:physical} and singleton service give
\begin{equation}
 \mathcal J_{N,s}(C)\ge
 \max\left\{1,\frac{C^2(N-s)}{4N^2(\Delta+C/(2N))}\right\}.
 \label{eq:capital-curve-lower}
\end{equation}
For $\Delta>0$, put $d=\lceil4(N-s)/\Delta\rceil$. If $d<N$, the all-size bipartite Ramanujan result~\cite{MSS2018} gives a regular multiplicity matrix with $\sigma^2\le4/d$. Its required working capital is at most $C$. Increasing every account to $C/N$ preserves service, and the Boolean support has at most $d$ neighbors per type. If $d\ge N$, use complete access instead. Thus
\begin{equation}
 \mathcal J_{N,s}(C)\le\frac CN
 \min\left\{N,\left\lceil\frac{4(N-s)}\Delta\right\rceil\right\}.
 \label{eq:capital-curve-upper}
\end{equation}
The degree is at least four because $\Delta\le N-s$. This upper construction uses the support degree only as an upper bound; it does not assert Boolean regularity.

For all sufficiently large $N$ and $0<\theta_-\le s/N\le\theta_+<1$, the two bounds imply, uniformly over $0\le\Delta\le N-s$,
\begin{equation}
 \mathcal J_{N,s}(s+\Delta)=\Theta\left(\frac N{\Delta+1}\right).
 \label{eq:capital-curve-order}
\end{equation}
In the near-efficient branch, the numerator of~\eqref{eq:capital-curve-lower} is bounded below by a constant times $N$, after division by $N^2$, and its remaining denominator by a constant times $\Delta+1$. Outside that branch $\Delta\ge s=\Theta(N)$, so singleton service supplies the constant lower bound. For $\Delta\ge1$,~\eqref{eq:capital-curve-upper} supplies the upper order; for $0\le\Delta<1$, complete access gives $O(N)$. Constants depend on the service-fraction bounds. This reciprocal curve explains the square-root balance under positive capital prices but does not give a tight finite approximation ratio over graphs.

\subsection{An exact endpoint at minimum working capital}
\label{app:minimum-working}
For integer $1\le s<N$,
\begin{equation}
 \mathcal J_{N,s}(s)=\frac{s(N-s+1)}N.
 \label{eq:minimum-working}
\end{equation}
At $C=s$, every positive-capital account must be accessible to at least $N-s+1$ types. Otherwise $s$ types could request all $s$ units while avoiding that account. Consequently
\[
 NJ\ge\sum_fa_fd_f\ge(N-s+1)C=s(N-s+1).
\]
For attainment, label types and accounts cyclically, let type $i$ access $i,i+1,\ldots,i+N-s$ modulo $N$, and put $s/N$ in every account. Repeated union with the next cyclic shift increases the size of a nonempty set by at least one until it reaches the whole cycle. Therefore
\[
 |N_H(U)|\ge\min\{N,N-s+|U|\}.
\]
For $|U|\le s$, multiplying by $s/N$ gives at least $|U|$, because the difference is $(s-|U|)(N-s)/N$. For $|U|\ge s$, all accounts are reached. Service follows, and every source has exposure $s(N-s+1)/N$.

At $N=57,s=16$ this gives $C=16$ and $J=224/19\simeq11.789474$. It improves on complete access at the same minimum working commitment. The square-root result remains an order comparison; the endpoint calculation and fixed-graph certificate refine different parts of it.

\subsection{Exact capital on the order-seven incidence graph}
\label{app:exact57}
We prove Claim~\ref{cl:exact57} using a finite nonexistence check. The graph, source-loss model, and service set are fixed: point--line incidence in $\operatorname{PG}(2,7)$, unit account ceilings, $s=16$, one full-account source failure, and common available junior capital. Both account balances and their total may vary. The result does not optimize the permission graph.

\paragraph{A smaller obstruction.}
Suppose sixteen types reached at most 46 of the 57 accounts. Choose eleven accounts avoided by those types. Their incident types belong to the other 41 types. The incidence matrix in our normalized indexing is symmetric, so exchanging its two sides gives an eleven-type set with at most 41 account neighbors. It therefore suffices to rule out that smaller set.

For eleven types, the moment bound~\eqref{eq:integer-neighbors} requires at least 41 neighbors. If exactly 41 are reached, their positive incidence counts satisfy
\[
 \begin{gathered}
 \sum m_f=88,\qquad \sum m_f^2=198,\\
 \sum_{m_f>0}(m_f-2)(m_f-3)=198-5\cdot88+6\cdot41=4.
 \end{gathered}
\]
Each summand is nonnegative. Counts of five or more are impossible. Writing $n_k$ for the number of accounts with count $k$, we obtain $n_1+n_4=2$. The account count and first moment then give
\[
 n_1+n_2+n_3+n_4=41,\qquad
 n_1+2n_2+3n_3+4n_4=88,\qquad
 n_3=8-3n_4\ge2.
\]
Thus at least one line contains exactly three selected points. The eleven-point set also contains a point off that line.

\paragraph{Normalizing the search.}
Any three distinct points on a line, together with one off the line, can be mapped to
\[
 (0,0,1),\quad(0,1,0),\quad(0,1,1),\quad(1,0,0)
\]
by an incidence-preserving projective transformation. To see this directly, choose representatives $p_1,p_2$ for two points on the line and $p_4$ for the off-line point. They form a basis. A representative of the third point has the form $p_3=a p_1+b p_2$ with $a,b\ne0$. Send $p_1$ to $a^{-1}(0,0,1)$, $p_2$ to $b^{-1}(0,1,0)$, and $p_4$ to $(1,0,0)$. This invertible map sends $p_3$ to $(0,1,1)$; the inverse transpose acts on line coordinates and preserves dot products.

The normalized line is $x=0$, and exactly three of its points are selected. The other seven points must be chosen off that line, excluding the fixed fourth point. There are $49-1=48$ candidates. This restriction is exhaustive for the putative obstruction, not a heuristic assumption about its shape.

\paragraph{Exhaustive test.}
The artifact represents each point's account neighborhood by a 57-bit integer. Starting with the four fixed points, it enumerates all increasing seven-point choices from the candidate list. A partial branch is pruned only if the union already has more than 41 accounts. Adding points cannot reduce the union, so pruning preserves every possible obstruction. The recursive step is:
\begin{quote}\ttfamily\small
search(start, needed, reached):\\
\hspace*{1em}if popcount(reached) > target: return false\\
\hspace*{1em}if needed == 0: return true\\
\hspace*{1em}for k = start, ..., len(candidates) - needed:\\
\hspace*{2em}if search(k+1, needed-1, reached OR mask[k]):\\
\hspace*{3em}return true\\
\hspace*{1em}return false
\end{quote}
For target 41, forward enumeration visits 1,009,410 partial states, prunes 883,031, and exhausts the search space without a solution. Reversing the candidate order visits 1,284,612 states and also exhausts the search space. With target 42 the same search returns an eleven-point set; its neighborhood is checked directly to have size 42. A separately coded C++ implementation agrees with the Python forward traversal. The reduction, monotone pruning rule, and completed search establish that every sixteen-type set reaches at least 47 accounts.

Two commands run the checks without network access:
\begin{quote}\ttfamily\small
python exact57/verify\_exact\_57.py\\
sh exact57/check\_cpp.sh
\end{quote} The Python program independently reconstructs the incidence matrix, compares it with \texttt{data/permissions\_57.csv}, verifies all Gram entries, runs both enumeration orders and the positive control, and checks the lower cuts. The C++ command compiles locally and uses the same exported neighborhood masks. Neither search relies on a floating-point solver, time limit, or reported optimality gap. Recorded counts are partial search states, not the number of all unpruned subsets.

\paragraph{Attainment and necessity.}
For $u=1,\ldots,15$, the integer bounds in Appendix~\ref{app:moment57}, multiplied by $16/47$, supply at least $u$ units. At $u=16$ the strengthened bound supplies exactly sixteen. Larger sets contain a sixteen-type subset. Thus every service cut holds at $a_f=16/47$, with
\[
 C=57\frac{16}{47}=\frac{912}{47},\qquad
 J=8\frac{16}{47}=\frac{128}{47}.
\]
The 57 lower cuts in Appendix~\ref{app:orbit57} give $47C\ge912$ for every serviceable allocation, and degree averaging gives $J\ge8C/57$. The uniform allocation attains both bounds. The moment allocation $8/23$ uses exactly $47/46$ times each of these minima; its separate certificate requires no exhaustive search.

\section{Actual supplier commitments and convexification}
\label{app:funding}
The actual funding function in Proposition~\ref{prop:frontier} may have kinks and depend jointly on working and junior quantities. All quantities required on the comparison domain must be available. This appendix derives the posted-price example and its approximation error; it is not an equilibrium supply model.

\begin{table}[ht]
\centering
\caption{Actual and relaxed procurement of the same 48 units of claim-paying capital. The quotes are hypothetical all-in payments for one period.}
\label{tab:supply}
\begin{tabular}{lrrrr}
\toprule
Calculation & Junior $j$ & Senior $v$ & Bill & Admissible?\\
\midrule
Convex relaxation & 32 & 16 & 8.00 & No\\
Actual procurement & 40 & 8 & 9.28 & Yes\\
\bottomrule
\end{tabular}
\end{table}

\subsection{The contract and its optimal split}
The schedule stated in the main text is
\begin{equation}
 \Phi(X)=r_{\rm j}X-(r_{\rm j}-r_{\rm s})
 \min\{\bar v,(X-b_{\rm att})_+\}.
 \label{eq:funding-schedule}
\end{equation}

For this derivation write $a=r_{\rm j}$, $b=r_{\rm s}$, and $q=b_{\rm att}$; $q$ here is the attachment, not the projective order. Let internal junior supply be sufficient on $0\le X\le X_{\max}$, with $X_{\max}\ge q+\bar v$. Junior units cost $a>0$; senior units cost $b\in(0,a)$, are capped at $\bar v$, and require junior amount at least $q$ whenever senior supply is positive. At target $X=j+v$, minimizing $aj+bv=aX-(a-b)v$ means maximizing admissible $v$. The maximum is $\min\{\bar v,(X-q)_+\}$, proving~\eqref{eq:funding-schedule}. The slopes are $a,b,a$, so
\[
 b(X_2-X_1)\le\Phi(X_2)-\Phi(X_1)\le a(X_2-X_1)
 \quad(0\le X_1\le X_2\le X_{\max}).
\]
With unlimited senior supply the final upward slope change is absent and the schedule is concave. With finite supply it is concave only until exhaustion; an interval containing both changes has neither global curvature sign.

Two junior suppliers, each with capacity 50 and quote $0.22j_i$, and a senior supplier with quote $0.06v$ and capacity 20 implement the numerical case on $[0,100]$. The junior amounts sum to $j$. For a recovered claim $D$, their aggregate principal loss is $\min\{D,j\}$, senior loss is $\min\{(D-j)_+,v\}$, and outside-user residual is $(D-X)_+$. Internal priority changes the provider split, not the total capital paid to those users.

The illustrative law $D\in\{0,40,100\}$ with probabilities $(.90,.08,.02)$ is compatible with a five-percent senior impairment-frequency mandate requiring $j\ge40$. Expected junior principal loss is at most $0.10j_i$ for each proportional supplier, while eligible senior expected loss is at most $0.02v$. The posted quotes cover these bounds and reservation payments of $0.12j_i$ and $0.04v$. They are fixed all-in quotes, and can exceed minimal expected-loss-based compensation. No second expected-claim charge is added to $aj+bv$.

\subsection{The exact convex envelope}
Participation is represented by
\[
 0\le v\le\bar v y,\qquad j\ge qy,\qquad y\in\{0,1\}.
\]
Relaxing $y$ to $[0,1]$ yields $0\le v\le\bar v$ and $v\le(\bar v/q)j$. At fixed $X=j+v$, the largest relaxed senior commitment is
\[
 \min\left\{\bar v,\frac{\bar v}{q+\bar v}X\right\}.
\]
Write $c_*=(aq+b\bar v)/(q+\bar v)$. The relaxed cost is
\[
 \widehat\Phi(X)=
 \begin{cases}
 c_*X,&0\le X\le q+\bar v,\\
 aX-(a-b)\bar v,&q+\bar v\le X\le X_{\max}.
 \end{cases}
\]
Its slopes are $c_*<a$, so it is convex. Below $q$, the gap from $\Phi$ is $(a-c_*)X$; between $q$ and $q+\bar v$, it is $(c_*-b)(q+\bar v-X)$; above $q+\bar v$, it is zero. Thus
\begin{equation}
 0\le\Phi(X)-\widehat\Phi(X)\le
 \delta_{\rm quote}:=\frac{(a-b)q\bar v}{q+\bar v},
 \label{eq:envelopegap}
\end{equation}
with equality at $X=q$. Any convex function below $\Phi$ lies below the chord from $(0,0)$ to $(q+\bar v,aq+b\bar v)$, and therefore below $\widehat\Phi$ on that interval. Beyond it, $\widehat\Phi=\Phi$. This proves that it is the greatest convex minorant.

For the numerical quotes, $c_*=1/6$, $\delta_{\rm quote}=32/15$, and the envelope is $X/6$ until 60, then $0.22X-3.2$. The actual cost at 48 is $8.8+.06(48-40)=9.28$. A convex average is not necessarily an admissible commitment: $(32,16)=.8(40,20)+.2(0,0)$ averages an active contract costing ten with a zero contract. It meets an expected budget of eight, not a hard budget of eight in every realization.

\subsection{What the gap means at a fixed budget}
Let $X(B)$ and $\widehat X(B)$ be the actual and relaxed affordable totals, capped at $X_{\max}$. Equation~\eqref{eq:envelopegap} shows that at most $\delta_{\rm quote}$ extra payment buys the relaxed target under the actual contract. With no budget increase, set $B_*=aq+b\bar v=c_*(q+\bar v)$. Before $B_*$,
\[
 X(B)=
 \begin{cases}B/a,&0\le B\le aq,\\
 q+(B-aq)/b,&aq\le B\le B_*,\end{cases}
 \qquad \widehat X(B)=B/c_*.
\]
The gap increases on $[0,aq]$ and decreases on $[aq,B_*]$; afterward the schedules coincide. Therefore
\[
 0\le\widehat X(B)-X(B)\le\delta_{\rm quote}/c_*.
\]
In the numerical example this maximum is $64/5=12.8$ at $B=8.8$. At $B=8$, actual capital is $400/11$, relaxed capital is 48, and the gap is $128/11$. Since $(D-X)_+$ is nonincreasing and one-Lipschitz in $X$, this is also a uniform upper allowance for the additional residual at any fixed claim $D$.

The map $X=j_1+j_2+v$ has rank one but does not preserve senior eligibility. Both $(j,v)=(40,0)$ and $(20,20)$ total 40, while only the first is admissible. Retaining $(X,j)$ uses two independent aggregates. Any broader supplier-level convexification must keep those participation constraints or supply a feasible repair. Standard convexification and pricing results apply under their own assumptions~\cite{HuaBaldick2017,UdellBoyd2016}; the explicit calculation here does not need a many-supplier theorem.

\subsection{Finite cost guarantees}
For the moment pair $S=(456/23,64/23)$, compare complete access $P=(16,16)$ through $(16,64/23)$:
\begin{align*}
 F(P)-F(S)&=F(16,16)-F(16,64/23)\\
 &\quad-[F(456/23,64/23)-F(16,64/23)]\\
 &\ge\frac{304}{23}\underline r_J-\frac{88}{23}\overline r_C.
\end{align*}
Comparing isolation $I=(57,1)$ through $(57,64/23)$ gives
\[
 F(I)-F(S)\ge\frac{855}{23}\underline r_C-\frac{41}{23}\overline r_J.
\]
Every difference changes one coordinate, so these comparisons permit a nonseparable bill. At working-price bounds $[.04,.08]$ and junior-price bounds $[.12,.30]$, the guarantees are $32/25=1.28$ against complete access and $219/230\simeq0.952174$ against isolation.

The spectral pair $S_0=(1311/64,23/8)$ provides a second sufficient commitment. Its guarantees are $13.125\underline r_J-4.484375\overline r_C=1.21625$ and $36.515625\underline r_C-1.875\overline r_J=0.898125$ at the same bands. Both allocations are evaluated on the same saved funding schedules; their result files identify the allocation explicitly.

For the moment pair, linear working/junior prices $(1,30)$ favor isolation, $(10,1)$ favor complete access, and $(.06,.20)$ favor the sparse candidate. Thus the design ranking depends on prices. Replacing common protection with dedicated protection changes the requirement to $J=C$: the moment design then needs $912/23\simeq39.652174$ combined units, while the spectral design needs $40.96875$.

In the square case, retaining sharper constants in the main proof gives
\[
 K^0(\Delta+\theta)\ge\theta^2(N-s)/4,\qquad\theta=s/N,
\]
and the funding lower bound $\theta\sqrt{\underline r_C\underline r_J(N-s)}-\underline r_C\theta$.

\subsection{Exact price regions for the three square candidates}
\label{app:prices}
Take the projective sufficient allocation with $N=q^2+q+1$ and $1<s<N$. Its excess working amount and reserve are
\[
 \Delta_S=\frac{q(N-s)}{(q+1)^2},\qquad
 J_S=\frac{q+s}{q+1}.
\]
At linear price ratio $\alpha=r_J/r_C>0$, let $E_I=N-s+\alpha$, $E_S=\Delta_S+\alpha J_S$, and $E_P=\alpha s$ be extra bills divided by $r_C$. Solving $E_S<E_P$ and $E_S<E_I$ gives
\[
 \alpha_-:=\frac{N-s}{(q+1)(s-1)}<\alpha<
 \alpha_+:=\frac{N(N-s)}{(q+1)(s-1)}.
\]
The two endpoints intersect at $\alpha=(N-s)/(s-1)$, between these thresholds. Hence complete access is cheapest below $\alpha_-$, sparse incidence between them, and isolation above $\alpha_+$, with ties at the two boundaries. At $N=57,s=16$, they are $41/120$ and $779/40$. At $s=1$, complete access has $(C,J)=(1,1)$ and dominates the other two candidates, so the threshold formula is not used. This is a price comparison among specified candidates, not a global finite optimum.

For the moment 57-account pair, the normalized bills are
\[
 E_I=41+\alpha,\qquad E_S=\frac{88}{23}+\frac{64}{23}\alpha,
 \qquad E_P=16\alpha.
\]
The equations $E_S=E_P$ and $E_S=E_I$ give $\alpha_-=11/38$ and $\alpha_+=855/41$. Complete access wins below the first threshold, the moment pair between them, and isolation above the second. These thresholds replace $41/120$ and $779/40$ only for the moment allocation at $s=16$; the general spectral formula above remains unchanged.

The minimum-working cyclic design in Appendix~\ref{app:minimum-working} has normalized bill $(224/19)\alpha$. Comparing it with the moment sparse pair gives $E_S<(224/19)\alpha$ exactly when $\alpha>209/492$. It can therefore beat the three-candidate winner at some prices.

\subsection{Moment allocation and separate capital prices}
\label{app:refined-prices}
Table~\ref{tab:capital} compares full protection of the same sixteen-unit service. The moment allocation lowers both commitments on the same sparse graph. It still uses more junior capital than isolation, while requiring much less working capital.
\begin{table}[t]
\centering\small
\setlength{\tabcolsep}{4.5pt}
\caption{Same service and common protection at $N=M=57$, $s=16$. The last column is a capital quantity; prices determine the funding bill. The three sparse rows fund the same permission graph using different service certificates.}
\label{tab:capital}
\begin{tabular}{@{}lrrrr@{}}
\toprule
Design & $C$ & $C-s$ & $J$ & $C+J$\\
\midrule
Isolated & 57 & 41 & 1 & 58\\
Complete access & 16 & 0 & 16 & 32\\
Sparse, spectral & 20.484375 & 4.484375 & 2.875000 & 23.359375\\
Sparse, moment & \RefinedWorking{} & \RefinedDelta{} & \RefinedJunior{} & \RefinedTotal{}\\
Sparse, exact & \ExactWorking{} & \ExactDelta{} & \ExactJunior{} & \ExactTotal{}\\
\bottomrule
\end{tabular}
\end{table}

Let $S=(456/23,64/23)$, $P=(16,16)$, and $I=(57,1)$. Coordinatewise application of the price bounds gives
\begin{align}
 F(P)-F(S)&\ge\frac{304}{23}\underline r_J-\frac{88}{23}\overline r_C,\notag\\
 F(I)-F(S)&\ge\frac{855}{23}\underline r_C-\frac{41}{23}\overline r_J.
 \label{eq:finiteprices}
\end{align}
For working prices in $[.04,.08]$ and junior prices in $[.12,.30]$, these guarantee savings of at least $1.28$ and $219/230\simeq0.952174$. The bounds allow a nonlinear, nonseparable bill. With linear relative price $\alpha=r_J/r_C$, the moment candidate is cheapest among these three when
\[
 \frac{11}{38}<\alpha<\frac{855}{41}.
\]
Complete access wins below that interval and isolation above it, with ties at the boundaries. Appendix~\ref{app:prices} retains the spectral comparison and the exact derivations. A cyclic design improves on complete access at $C=s$ (Appendix~\ref{app:minimum-working}); neither endpoint comparison asserts a global finite optimum.

Dedicated junior payment rights instead require $J=C$, making the moment candidate's combined capital $912/23\simeq39.652174$, greater than complete access's 32 under common protection. Operating costs can also change the ranking. The funding objective does not price the execution, custody, or maintenance of the permission relation.

\subsection{Prices at the exact fixed-graph optimum}
\label{app:exact57-prices}
Let $S_*=(912/47,128/47)$ be the capital pair in Claim~\ref{cl:exact57}. The lower bounds on its two coordinates hold for every serviceable allocation on this graph and are attained together. Hence $S_*$ minimizes any coordinatewise increasing funding bill on this fixed graph. This conclusion does not compare it with every other permission graph.

Comparing $S_*$ to $P=(16,16)$ and $I=(57,1)$ along one coordinate at a time gives
\[
 F(P)-F(S_*)\ge\frac{624}{47}\underline r_J-\frac{160}{47}\overline r_C,
 \qquad
 F(I)-F(S_*)\ge\frac{1767}{47}\underline r_C-\frac{81}{47}\overline r_J.
\]
Under the working-price bands $[.04,.08]$ and junior-price bands $[.12,.30]$, these are at least $1552/1175\simeq1.320851$ and $2319/2350\simeq0.986809$, respectively. For linear prices, direct comparison gives
\[
 \frac{10}{39}<\alpha<\frac{589}{27}
\]
as the interval in which $S_*$ is cheaper than both isolation and complete access. At the boundaries the relevant candidates tie. The minimum-working cyclic pair $(16,224/19)$ is cheaper than $S_*$ for $\alpha<95/253$, and more expensive for $\alpha>95/253$. This comparison uses that particular cyclic pair, not a claim that these candidates exhaust the global frontier.

Dedicated payment rights require $J=C$ and give combined capital $1824/47$. Under common cover and the illustrative $\alpha=3$, $\pi=1/10$ convention of Appendix~\ref{app:benchmark}, the extra bill divided by $r_C$ is $544/47$. Removing the constant benchmark component gives $5299/470$. These are funding calculations under assumed prices, not observed insurance quotes.

\section{Conditions on repeated borrowing}
\label{app:continuation}
Proposition~\ref{prop:continuation} uses a bound on total gross claims together with a feasible full-protection policy. Two ways to obtain the amount bound follow. In either case, the controls must preserve ordinary service and the accepted coverage, and the policy must finance every payment from the capital then available. A bound on the unpaid residual of a partial stabilizer is insufficient.

\begin{proof}[of Proposition~\ref{prop:continuation}]
The original loss still requires $J\ge K^0$, preserving the static lower bound. The stipulated sparse policy needs at most $\min\{M,\Gamma K^0\}=O(\sqrt N)$ common junior capital, so its working capital, protection, and allowed control costs give the upper bound.

For unrestricted reach, let component $h$ contain $n_h$ types and working capital $C_h$. Service requires $C_h\ge\min\{n_h,s\}$. If every component has fewer than $s$ types, then $C\ge\sum_hn_h=N$ and $C-s\ge N-s$. Otherwise a component contains at least $s$ types, and its permitted compromise requires $J\ge C_h\ge s$. Thus
\[
 F_{N,M}(C,J)-F_{N,M}(s,0)
 \ge\min\{\underline r_C(N-s),\underline r_Js\}=\Omega(N).
\]
Complete access with $C=s,J=s$ is feasible for $M\ge s$ and has $O(N)$ extra cost.
\end{proof}

\subsection{Two sufficient ways to control continuation}
\paragraph{A cumulative inherited allowance.}
Give all continuations descended from source $i$ one non-replenishing allowance of $\Gamma a(N_H(i))$, with $\Gamma\ge1$ independent of system size. Each new covered dollar consumes one unit of this allowance. Summing the debits bounds gross claims by $\Gamma K^0$. Every approved wrapper and account alias must retain the same source identity. Changing an identifier or repaying outstanding debt must not reopen the allowance. The borrowing policy must also continue to serve the requests promised for the coverage period.

\paragraph{A finite number of stages and bounded monetary branching.}
Suppose the first new loss is at most $K^0$, at most $h$ further waves occur, and each creates no more than $g$ times the preceding wave's new covered amount. Induction bounds wave $t$ by $K^0g^t$, so total gross loss is at most
\[
 K^0\sum_{t=0}^{h}g^t.
\]
Fixed $h$ and $g$ make the factor independent of system size. Both restrictions are needed: a bound on path length alone allows a single step to create claims at arbitrarily many accounts. If timely payments are what enforce the branching bound, their affordability must be checked after previous payments have reduced the reserve. Principal limits and one-time recognition of each covered liability still apply.

The projective graph illustrates the opposite case. Every pair of types shares an account, so the inheritance rule in Section~\ref{sec:continuation} reaches all types after one additional type-to-type step and then reaches all accounts. Each newly reached balance can be drawn only once. Refusing those inherited permissions or enforcing persistent source-origin limits can stop component exhaustion, thereby removing the negative proposition's premise.

\section{Experimental definitions and reproducibility}
\label{app:evidence}
The experiments below test the formulas on finite inputs. We report exact rational values separately from floating-point residuals and retain the counterexamples to stronger interpretations, including increased exposure after account reassignment.

\subsection{Static experiments}
The square service experiment enumerates all Boolean matrices for $N=2,3$ and all integer $1\le s<N$. For each serviceable matrix it computes the minimum uniform amount using all service cuts and independently solves the extremal request flows. The heterogeneous family uses random square graphs with diagonal service options, optimized capacities, and nonnegative perturbations within unit account limits. The near-efficient predicate is evaluated before applying Proposition~\ref{prop:physical}. Separate retained shortfall checks use the variant in Appendix~\ref{app:construction}.

Prime incidence matrices are constructed by normalized triples and tested through exact integer multiplication. The exact incidence certificate is checked independently of the numerical flow solutions. Its service tests select extremal unit requests under the declared total bound; the supplementary quota checks also include fractional requests. Common and dedicated capital placements settle the same source claims.

The actual cost family is
\[
 F_N(C,J)=f_C(C)+f_J(J)
   +\gamma N(1-e^{-C/N})(1-e^{-J/N}),
\]
where $f_C,f_J$ integrate piecewise-constant positive slopes. Working slopes lie in $[.04,.075]$, junior slopes in $[.12,.295]$, and $0\le\gamma\le.005$. Each interaction partial derivative lies in $[0,\gamma]$, proving the bands $[.04,.08]$ and $[.12,.30]$ globally on nonnegative quantities. We evaluate these functions directly, using fixed seeds and retaining every scalar output.

The rectangular pipeline uses prime orders two, three, five, and seven, three service fractions, and several actual account counts with $C\le M$. Each of its 59 combinations has six tested requests. The overlap-transferred flow and an independently optimized actual flow are both checked. Ninety irregular rectangular books start with $s$ fully connected columns to guarantee service; optimized capacities receive small positive perturbations within their unit bounds. Eighty-seven meet the near-efficient hypothesis. The exhaustive no-headroom test is over all matrices in the three stated small dimensions, not over a random sample.

Protection partitions assign account $f$ to pool $f\bmod p$ for the supplied pool counts. The exact sum of each pool's maximal assigned claim is compared with a separate capital LP. The chosen partitions are inputs to this comparison. Component-access tests enumerate graph components and compare the service and loss amounts in Proposition~\ref{prop:continuation}.

\subsection{Exact supplementary checks}
A separate rational suite verifies the numerical commitments in Table~\ref{tab:supply} and the spectral entries retained in Table~\ref{tab:capital}, the convex-envelope gap $32/15$, the affordable-capital gap $64/5$, and repairs at fixed capital and fixed budget. It checks the slope inequalities on every union breakpoint interval and tests the maximum-gap witnesses. These checks implement the formulas proved in Appendix~\ref{app:funding}.

For complete permissions and rational inputs with $s\le M$, the quota test checks $z_{if}=x_i/M$ against $L_{if}=1/M$. A source attains its unit total without violating an account guard. Once the cumulative limits are used, every further positive draw is rejected. The recorded operations are logical transitions; they supply no storage or gas estimate.

\subsection{Reproduction}
The artifact has separate commands for the static evaluation and the broader regression suite. Its results index maps each reported output to its inputs and generating program. Generated families use fixed seeds. NumPy and SciPy calculations run with one linear-algebra thread; the environment record and result files include package versions and maximum numerical residuals. Rational checks evaluate their witnesses directly, without accepting a solver's reported optimum as a certificate.

A second run from a separate working directory checks reproducibility by comparing SHA-256 output hashes. This comparison concerns the same inputs and program. The deployment and data limitations are collected in Section~\ref{sec:empirical}.

\subsection{Service, degree, and price checks}
The separate command \texttt{python -m src.security\_examples} checks the equal-service figure and degree certificates with exact fractions. Three designs each satisfy 127 nonempty service cuts, and 192 rational flow witnesses cover all requests with zero, one, two, or three unit demands. Thirty-five witnesses per design have total demand three. The sparse graph's minimum uniform amount is exactly $1/2$; its spectral $s=2$ witness is retained separately.

Sixty degree/price settings compare the rounded coarse certificate minimizer with exhaustive integer evaluation. The smaller factors retaining $(d-1)/d$ are checked against the coarse bounds. This test does not construct the all-size graph. The price-region calculation tests 55 service levels at $N=57$, including both boundary ties and a point in each of the three regions, for 275 exact comparisons. The collinear equal-split test records a routing failure at the spectral account commitment. The artifact index identifies these five result files.

\subsection{Underwriting computations}
The spectral, moment, and exact fixed-graph allocations use the same normalized order-seven incidence and the same saved 96 requests and 360 funding schedules. The input-export program preserves the generator's random-number sequence; the resulting fixture is bound by a SHA-256 digest. Each result file identifies the allocation being evaluated.

The moment checker verifies every Gram entry, the sixteen-type lower witness and its permutations, and account occurrences in all 57 service cuts. It checks the size-dependent integer bounds directly. The exact checker adds the obstruction reduction and complete searches described in Appendix~\ref{app:exact57}. A separate C++ program checks the normalized search using the exported incidence masks. Neither search uses a solver, a time limit, or a heuristic stopping condition.

For each of the moment and exact allocations, rational max flows serve all 96 saved requests; independent numerical LPs check the same requests. The general service guarantees come from the moment inequalities and exhaustive obstruction proof, respectively, not from the sampled requests. Seven fixed protection partitions compare direct assigned claims with independently optimized payment allocations. Explicit collinear subsets attain the one-, two-, four-, and eight-source losses.

The nonlinear evaluation retains every breakpoint, slope, and interaction coefficient. Its rows distinguish isolation, complete access, the spectral and moment certificates, the exact fixed-graph allocation, and the minimum-working cyclic design. Across the 360 schedules, the exact allocation's minimum savings are \ExactMinSavingIsolated{} against isolation, \ExactMinSavingComplete{} against complete access, and \ExactMinSavingMoment{} against the moment allocation. These calculations check the stated price-band implications; they do not estimate market funding terms.

The cyclic endpoint constructor is checked on 45 instances with $2\le N\le10$ and every integer $1\le s<N$, totaling 16,343 nonempty service cuts. Another 324 rational checks compare equivalent budget formulas. These finite tests accompany the general proofs in Appendix~\ref{app:underwriting}.

\subsection{Historical material and protocol scope}
The retained Kelp report-derived file contains conditional claim figures, not a price series or an estimated benchmark law. Euler's retrospective describes collateral deterioration and stale pricing during the late-2025 Stream crisis, while reporting no direct exposure in its DAO-managed markets~\cite{EulerRetrospective2026}. These accounts motivate the failure and configuration questions in Section~\ref{sec:security}. Neither establishes the full-account attack set used by the theorem. Historical price dislocations require separate instrument identities, event anchors, reference prices, and observations before they can be reproduced; converting them into claims additionally requires debt and recovery records.

\subsection{Notation used across the examples}
\begin{center}\small
\begin{tabular}{ll}
\toprule
Symbol & Meaning\\
\midrule
$N,M,s$ & Borrowing types, lending accounts, aggregate service bound\\
$H,a,C,\Delta$ & Permissions, working amounts, total working capital, $C-s$\\
$K^0,J$ & Pre-cover source loss and common outside junior capital\\
$J_H^*,\bar C,\bar J$ & Fixed-graph junior optimum and the two capital budgets\\
$\mathcal J_{N,s}(C)$ & Best junior amount over graphs at exactly $C$ working capital\\
$A,j,Q$ & Payment eligibility, pool balances, liability ownership\\
$d,\sigma$ & Spectral witness degree and normalized second singular value\\
$q$ & Projective order; Appendix~\ref{app:funding} uses $q=b_{\rm att}$\\
$q_i,Q,L$ & Empirical type caps, their sum, and their maximum (Section~\ref{sec:empirical})\\
$\kappa_i$ & Cumulative, non-replenishing source allowance\\
$r_C,r_J,\alpha$ & Working price, outside junior price, $r_J/r_C$\\
$r_{\rm j},r_{\rm s},b_{\rm att}$ & Internal provider quotes and senior-entry attachment\\
$\Gamma,M^0$ & Continuation factor and benchmark mean, respectively\\
\bottomrule
\end{tabular}
\end{center}

\section{Exact experiments and historical calibration}
\label{app:empirical}
The empirical evidence directory contains tables, saved calibrations, numerical witnesses, cached observations, source code, and SHA-256 records. Separate commands check exact certificates, reproduce numerical comparisons, and reconstruct calibrations where the required raw observations are available.

\subsection{Cohorts and fitting windows}
\label{app:cohorts}
The three calendar anchors are 9 January 2025 (Usual), 22 March 2026 (Resolv), and 18 April 2026 (Kelp). Usual's January announcement changed the exit arrangements for its long-dated USD0++ product~\cite{Usual2025}; the anchor is not classified here as a software exploit. Resolv's postmortem describes a signing-system compromise and unauthorized USR minting on its March anchor~\cite{ResolvPostmortem2026}. The Kelp anchor follows the bridge incident report~\cite{KelpReport2026}. They index the stored daily series and do not imply that every observation shares the precise onchain incident time. Table~\ref{tab:cohort-scope} identifies the modeled groups. For each cohort, the same selected markets enter every rolling date; selection uses the retrospectively acquired catalog and is not an ex ante sampling rule. The stored market IDs and raw-response hashes are in \texttt{hindsight\_ratios.json}.

\begin{table}[htbp]
\centering\small
\caption{Scope of the grouped Morpho cohorts. Cutoffs are UTC dates for the saved day-$-1$ witnesses. These five accounts are modeling aggregates, not five observed lender vaults.}
\label{tab:cohort-scope}
\begin{tabular}{@{}llr p{5.1cm}@{}}
\toprule
Cohort & Cutoff & Markets & Five source groups\\
\midrule
Usual & 2025-01-08 & 11 & BTC, DAI, ETH, Ethena, Usual \\
Resolv & 2026-03-21 & 32 & BTC, ETH, Ethena, Resolv, Usual \\
Kelp & 2026-04-17 & 15 & BTC, Etherfi, Kelp, Lido, Renzo \\
 
\bottomrule
\end{tabular}
\end{table}

At every daily cutoff $t$, the previous 28 dated observations, including $t$, give
\[
 q_{i,t}=\max_{u\text{ in window}}d_i(u),\qquad
 s_t=\max_{u\text{ in window}}\sum_i d_i(u).
\]
Group debt $d_i$ is outstanding principal in native units. Treating these stocks as independent batch requests is the counterfactual service definition. Accounts have ceilings $\widetilde b_{i,t}=\max\{b_{i,t},q_{i,t}\}$, where $b_{i,t}$ is observed group supply at $t$. Working allocations may be reassigned within this grouped model; this is not a refinancing or transaction replay of existing loans. All compared designs receive the same ceilings and source set at a date. The source attack exhausts each funded accessible account once, with zero recovery, and every design buys its own sufficient common junior capital.

\begin{figure}[p]
\centering
\includegraphics[width=\linewidth]{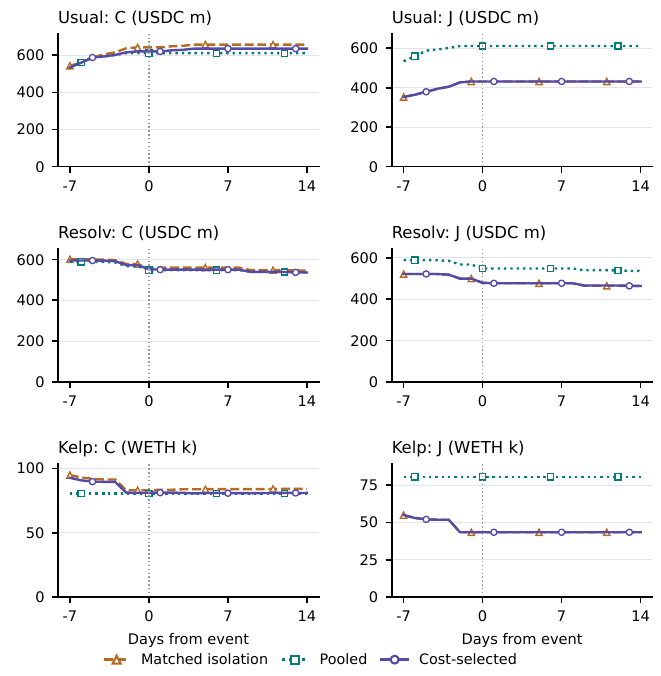}
\caption{Working capital and required junior capital for the three Morpho-derived cohorts. Each date independently uses its trailing fit and repaired capacity. Coincident junior curves often reflect the same worst-source requirement. Native USDC and WETH quantities are not combined into a cross-cohort dollar ranking.}
\label{fig:rolling-supplement}
\end{figure}

Figure~\ref{fig:rolling-supplement} shows all 22 independent redesigns. At each date the saved mixed-integer calculation minimizes $C+3J$ over every Boolean five-by-five graph and continuous balances. Its largest stored relative primal--bound gap is $5.05\times10^{-7}$. This is numerical global optimization of the fitted grouped model, not an exact rational certificate or a protocol-wide optimum. Re-solving can return another tied graph. The frozen diagnostic uses the saved witness, not a replacement selected for later performance.

\subsection{What capital has been added?}
Let $T_t=\sum_i(\widetilde b_{i,t}-b_{i,t})$ be added capacity and $A_t=\sum_i(a_{i,t}-b_{i,t})_+$ be the selected allocation above observed per-account supply. The first is an available ceiling, while the second counts actual commitment beyond observed account supplies. Table~\ref{tab:cutoff-capital} gives both at the fixed pre-event cutoff. Neither quantity estimates a historically executable supplier quote. Positive allocations, including additions, are counted in the funding objective.

\begin{table}[htbp]
\centering\small
\setlength{\tabcolsep}{4pt}
\caption{Pre-event capital adjustment in native units. Added ceilings and funded additions are distinct. All designs share the same adjusted ceilings.}
\label{tab:cutoff-capital}
\begin{tabular}{@{}lrrrr@{}}
\toprule
 & Observed & Added & Selected & Funded\\
Cohort (units) & supply & capacity $T_t$ & $C$ & additions $A_t$\\
\midrule
Usual (m USDC) & 724.763 & 39.423 & 620.699 & 39.423 \\
Resolv (m USDC) & 556.903 & 23.559 & 572.459 & 23.511 \\
Kelp (k WETH) & 74.471 & 8.568 & 81.110 & 6.652 \\
 
\bottomrule
\end{tabular}
\end{table}

Post-event changes can be larger than those pre-event additions. At Usual day $+12$, $T_t=488.009$ million USDC and $A_t=466.256$ million, against observed supply of 169.571 million and selected $C=635.828$ million. At Kelp day $+3$, $T_t=27{,}407.34$ WETH and $A_t=24{,}583.91$ WETH. The per-date audit retains these quantities for every design. A plateau in a rolling fit can result from an earlier debt maximum remaining in the window; it need not reflect a stable post-event funding position.

\subsection{Exact synthetic topology optimization}
\label{app:exact-experiments}
The deterministic suite has 27 cases with three types and accounts, and one with four. Account ceilings are identical within each case, so column permutations preserve feasibility and objective even when type demands differ. Enumeration covers 120 canonical column multisets for each three-type case and 3,876 for the four-type case, including repeated and zero columns. These represent all 512 and 65,536 labeled permission matrices, respectively.

A floating-point LP proposes each fixed-graph optimum. Rational reconstruction must then establish every service and attack constraint, primal bounds, dual signs, stationarity, and exact equality of the primal and dual objectives. An infeasible support needs a service cut that fails even at maximum account funding. Every published support is certified: 3,897 exact feasible optima and 3,219 infeasible cuts across 7,116 supports. Taking the least objective across supports certifies a global finite optimum. A solver failure does not count as an infeasible graph.

The suite varies service $s$, relative price $\alpha$, type imbalance, supply ceilings, source coalitions, hard cumulative loss controls, and permission-component continuation. Repeated anchor cases serve different sensitivity comparisons; they are not independent draws. Cases with an infeasible endpoint are retained but excluded from strict two-endpoint percentage comparisons. The full result table and case configurations are available in machine-readable form.

\begin{figure}[p]
\centering
\includegraphics[width=\linewidth]{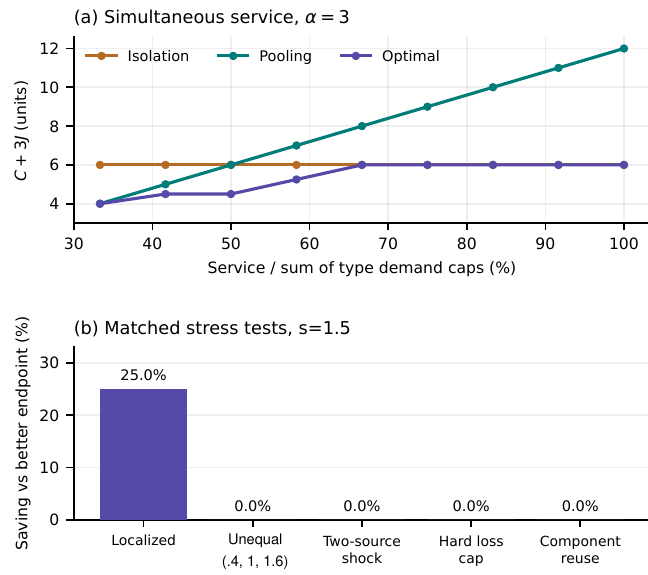}
\caption{Exact finite sensitivity results, with a freshly optimized graph in each case. The uneven-demand case uses $q=(.4,1,1.6)$, or $\beta=.6$. The other stresses change the source budget, cumulative loss cap, or component-reach rule. Their loss of a strict advantage concerns these three-type contracts, not a universal threshold for larger systems.}
\label{fig:synthetic-stress}
\end{figure}

For $q=(1,1,1)$ and $\alpha=3$, strict savings occur at sampled $s=1.25,1.5,1.75$; complete access is optimal at $s=1$, and isolation is optimal from sampled $s=2$ onward. The unequal-demand family uses $q=(1-\beta,1,1+\beta)$ and identical ceilings $1+\beta$. Its strict savings shrink as the largest source grows. Coalition size is a worst-case source budget, not a statistical correlation. Component continuation closes the permission graph even through currently empty accounts; it does not model liquidity refills or transaction feasibility beyond that specified closure.

\subsection{Frozen saved-witness service diagnostic}
For each cohort, fix the day-$-1$ graph, balances, and endpoint allocations exactly as stored. Recover the fifteen daily debt vectors for days 0--14 from the overlapping calibrated windows, checking that repeated observations agree. The shortage for one vector is
\[
 \operatorname{shortfall}(H,a,x)
 =\max\left\{0,\max_{U\subseteq[5]}\{x(U)-a(N_H(U))\}\right\}.
\]
No account principal carries over between those independent feasibility tests. Enumeration of all 31 nonempty cuts is checked against an independent max-flow LP for each of 135 design/day pairs. In the recorded run the largest discrepancy is below $6\times10^{-8}$ native units. The served criterion is shortage at most $10^{-6}$ model units, corresponding to one USDC or .001 WETH. Maximum shortages accompany counts; Kelp isolation's first shortage is only .00218 WETH.

All failed vectors exceed at least one frozen type cap; none exceeds its frozen total $s$. Resolv and Kelp's saved cost-selected designs serve all fifteen vectors, while their isolated witnesses do not. Usual serves seven under either selected or isolated access. Complete access serves all fifteen for each cohort. This is a retrospective diagnostic of saved witnesses. It uses no holdout-based tie-break, does not reoptimize, and does not establish stability of all tied optima or prospective performance under another sample. The cohorts were still chosen retrospectively.

\subsection{Benchmark-adjusted loss and source-specific exposure}
\label{app:adp-diagnostic}
For an illustrative primitive source law, with probability $1-p$ there is no failure. Otherwise, one of the five groups is chosen uniformly and exhausts its permissions. At $p=.1$,
\[
 \ell=Ha,\qquad K^0=\max_i\ell_i,\qquad
 M^0=\frac{p}{5}\sum_i\ell_i,\qquad \ADP^0=K^0-M^0.
\]
The law is common, but the mean changes with the architecture. Singleton service forces $\ell_i\ge q_i$ in these fits; isolation attains equality. At unchanged $K^0$, any ADP reduction therefore comes from a higher mean. Among 66 windows, 58 have unchanged $K^0$ to tolerance and 56 of those have lower ADP. All eight windows with higher selected $K^0$ also have lower ADP. These changes cannot establish reduced required protection.

For Kelp's date-zero fit, isolation and the selected graph both have $K^0=43{,}423.54$ WETH. Their assumed means are 1,659.72 and 3,921.06 WETH, giving ADPs 41,763.82 and 39,502.48. The 2,261.34-WETH difference is entirely the mean change. All designs have zero residual ADP once fully protected. The artifact retains the full losses, mean, and difference for $p\in\{0,.01,.1,.5,1\}$; no probability or insurance yield is estimated.

The max-source objective also allows nonbinding sources to become more exposed. In the frozen Kelp fit, modeled Kelp-group loss grows from 496.19 to 37,743.09 WETH while the Lido-group maximum stays at 43,423.54. In Resolv, the named group grows from 5.199 million to 73.223 million USDC while the BTC-group maximum stays fixed. These are modeled source-exhaustion amounts, not incident loss reconstructions or prevented-loss estimates.

\subsection{Reproduction boundary}
Reproduction separates three tasks. Exact synthetic checking uses all exported rational certificates. Saved-input validation recomputes capital, service cuts, cost, conditional benchmark quantities, and the frozen diagnostic; finite historical models can also be numerically reoptimized. Acquisition reconstruction additionally requires the raw market histories and grouping configuration. All 22 Usual calibrations can be reconstructed from included API caches. Seven bound raw inputs for Resolv and Kelp are absent from the supplied bundle. The missing-input inventory is explicit; transformations from saved calibrations do not authenticate their acquisition.

The artifact contains the source snapshots used for these calculations and records the path and time adapters needed to run them independently of their parent repository. Those adapters do not change the economic functions. Reproduction scripts distinguish numerical solver bounds from exact rational certificates, generate the paper tables from the checked results, and preserve the saved witnesses for the frozen diagnostic. They send no transactions and acquire no new observations. No price-dislocation or realized-loss series is estimated.

\section{Protocol mapping and execution obligations}
\label{app:protocol}
\begin{table}[t]
\centering\small
\setlength{\tabcolsep}{4pt}
\caption{Candidate Euler mapping, not a deployed implementation. The middle column names documented mechanisms; the last gives obligations still to establish.}
\label{tab:mapping}
\begin{tabular}{@{}>{\raggedright\arraybackslash}p{2.0cm}>{\raggedright\arraybackslash}p{4.0cm}>{\raggedright\arraybackslash}p{5.4cm}@{}}
\toprule
Model object & Candidate mechanism & Required check\\
\midrule
Type and access & Collateral-vault relationships and controller checks~\cite{EulerMarkets} & Group common failures; include LTVs and all effective amount limits.\\
Working account & Lending vault or separately accounted commitment & Identify the covered principal and prevent duplicate use.\\
Observed allocation & EVC atomic batch~\cite{EulerEVC} & Produce a feasible allocation with sufficient collateral and valid final account checks.\\
Borrowing restriction & Selected-operation hook~\cite{EulerHooks} & Cover every relevant entry point and receipt path.\\
Junior protection & Additional funded payment contract & Bind claim eligibility, custody, availability, and one-time settlement.\\
\bottomrule
\end{tabular}
\end{table}
Euler's liability vaults specify collateral relationships, loan-to-value limits, oracle routes, and caps~\cite{EulerMarkets}. The EVC supports batched operations and deferred checks~\cite{EulerEVC}; an allocator must still supply a feasible funding assignment. At an account-status check, an EVC account has at most one controller. Multi-vault borrowing needs suitable subaccounts or reviewed coordinating logic, without pledging the same collateral independently in full to each~\cite{EVCWhitepaper}. One modeled source may span several accounts.

Hooks reject selected operations~\cite{EulerHooks}. A protection implementation must cover every relevant entry point, preserve the source and principal accounting, and add a funded junior-payment contract. Equation~\eqref{eq:execution} is enough for conservative cover. Applying the architecture lower bound to a deployment additionally requires attainable neighborhood loss and the stated service guarantee. None of the historical calibration experiments establishes these deployment obligations.

\end{document}